# Low Complexity Hybrid Precoding Designs for Multiuser mmWave/THz Ultra Massive MIMO Systems

João Pedro Pavia[1,3], Student Member, IEEE, Vasco Velez[1,3], Renato Ferreira[1,3], Nuno Souto[1,3], Senior Member IEEE, Marco Ribeiro[1,3], João Silva[1,3], and Rui Dinis[2,3] Senior Member IEEE

[1]Department of Information Science and Technology, ISCTE-Instituto Universitário de Lisboa,1649-026 Lisboa, Portugal
[2]Department of Electrical and Computer Engineering, Faculty of Science and Technology, Universidade Nova de Lisboa, 1099-085 Lisboa, Portugal
[3]Instituto de Telecomunicações, 1049 - 001 Lisboa, Portugal
Corresponding author: João Pedro Pavia (e-mail: jpavia@lx.it.pt).

*Abstract*— Millimeter-wave and terahertz technologies have been attracting attention from the wireless research community since they can offer large underutilized bandwidths which can enable the support of ultra-high-speed connections in future wireless communication systems. While the high signal attenuation occurring at these frequencies requires the adoption of very large (or the so-called ultra-massive) antenna arrays, in order to accomplish low complexity and low power consumption, hybrid analog/digital designs must be adopted. In this paper we present a hybrid design algorithm suitable for both mmWave and THz multiuser multiple-input multiple-output (MIMO) systems, which comprises separate computation steps for the digital precoder, analog precoder and multiuser interference mitigation. The design can also incorporate different analog architectures such as phase shifters, switches and inverters, antenna selection and so on. Furthermore, it is also applicable for different structures namely, fully connected, arrays of subarrays (AoSA) and dynamic arrays of subarrays (DAoSA), making it suitable for the support of ultra-massive MIMO (UM-MIMO) in severely hardware constrained THz systems. We will show that, by using the proposed approach, it is possible to achieve good trade-offs between spectral efficiency and simplified implementation, even as the number of users and data streams increases.)

## I. INTRODUCTION

Over the last few years, significant advances have been made to provide higher-speed connections to users in wireless networks with several novel technologies being proposed to achieve this objective. However, future generations of communication systems will have to fulfil more demanding requirements that cannot be met by the methods adopted in today's communications systems. This motivates the exploration of other candidate technologies like the millimeter wave (mmWave) and Terahertz (THz) bands. These bands offer great underutilized bandwidths and also allow a simplified implementation of large antenna arrays, which are crucial to combat the severe signal attenuation and path losses that occurs at these frequencies [1]-[4]. While these technologies (THz systems in particular), are expected to ease the spectrum limitations of today's systems, they face several issues, such as the reflection and scattering losses through the transmission path, the high dependency between distance and frequency of channels at the THz band and the need of controllable time-delay phase shifters, since the phase shift will vary with frequencies based on the signal traveling time, which will also affect the system performance. These limitations require not only the proper system design, but also the definition of a set of strategies to enable communications [5], [6].

The exploration of the potentialities of millimeter and sub-millimeter wavelengths is closely related to the paradigm of using very large arrays of antennas in beamforming architectures. This gives rise to the so-called ultra-massive multiple-input multiple-output (UM-MIMO) systems. Still, to achieve the maximum potential of these systems it is necessary to consider the requirements and the challenges related not only to the channel characteristics but also to the hardware component specially regarding THz circuits [5], [7], [8]. Considering that high complexity and power usage are pointed out as the major constraints of large-antenna systems, the adoption of hybrid digital-analog architectures becomes crucial to overcome these issues. By adopting this type of design, it is possible to split the signal processing into two separate parts, digital and analog, and obtain a reduction of the overall circuit complexity and power consumption [9]. Adopting a proper problem formulation, the analog design part can then be reduced to a simple projection operation in a flexible precoding or combining algorithm that can cope with different architectures, as we proposed in [10], [11]. Despite the ultra-wide bandwidths available at mmWave and THz bands, and besides considering the problem of distance limitation, MIMO systems should take into account the operation in frequency selective channels [12]. To make the development of hybrid schemes for these systems a reality, it is necessary to handle the fading caused by multiple propagation paths typical in this type of channels [13]. Therefore, solutions inspired on multi-carrier schemes, such as orthogonal frequency division multiplexing (OFDM) are often adopted to address such problems [14].

Spectral Efficiency (SE) of point-to-point transmissions is a major concern in SingleUser (SU) and MultiUser (MU) systems. To achieve good performances, it is necessary to develop algorithms that are especially tailored to the architecture of these systems. Several hybrid precoding schemes have been proposed in the literature [16]-[18]. The authors of [15] proposed two algorithms for low complexity hybrid precoding and beamforming for MU mmWave systems. Even though, they assume only one stream per user, i.e., the number of data streams ($N_s$) is equal to the number of users ($N_u$), it is shown that the algorithms achieve interesting results when compared to the fully-digital solution. The concept of precoding based on adaptive RF-chain-to-antenna was introduced in [16] for SU scenarios only but with promising results. In [17], a nonlinear hybrid transceiver design relying on Tomlinson-Harashima precoding was proposed. Their approach considers fully-connected architectures only but can achieve a performance close to the fully-digital transceiver. A Kalman based Hybrid Precoding method was proposed for MU scenarios in [18]. While designed for systems with only one stream per user and based on fully connected structures, the performance of the algorithm is competitive with other existing solutions. A hybrid MMSE-based precoder and combiner design with low complexity was proposed in [19]. The algorithm is designed for MU-MIMO systems in narrowband channels, and it presents lower complexity and better results when compared to Kalman's precoding. Most of the hybrid solutions for mmWave systems aim to achieve near-optimal performance using Fully-Connected (FC) structures, resorting to phase shifters or switches. However, the difficulty of handling the hardware constraint imposed by the analog phase shifters or by switches in the THz band is an issue that limits the expected performance in terms of SE.

Array-of-SubArrays (AoSAs) structures have gained particular attention over the last few years as a more practical alternative to FC structures, especially for the THz band. In contrast to FC structures, in which every RF chain is connected to all antennas via an individual group of phase shifters (prohibitive for higher frequencies), the AoSA approach allows us to have each RF chain connected to only a reduced subset of antennas. The adoption of a disjoint structure with fewer phase shifters reduces the system complexity, the power consumption and the signal power loss. Moreover, all the signal processing can be easily carried out at the subarray level by using an adequate number of antennas [6].

Following the AoSA approach, it was shown in [20] that, to balance SE and power consumption in THz communications, adaption and dynamic control capabilities should be included in the hybrid precoding design. Therefore, Dynamic Arrays-of-SubArrays (DAoSA) architectures could be adopted. The same authors proposed a DAoSA hybrid precoding architecture which can intelligently adjust the connections between RF chains and subarrays through a network of switches. Their results showed that it is possible to achieve a good trade-off for the balancing between the SE and power consumption.

Within the context of multiuser downlink scenarios, the authors of [21] studied some precoding schemes considering THz massive MIMO systems for Beyond 5$^{th}$ Generation (B5G) networks. Besides showing the impact on EE and SE performance, carrier frequency, bandwidth and antenna gains, three different precoding schemes were evaluated and compared. It was observed that the hybrid precoding approach with baseband Zero Forcing for multiuser interference mitigation (HYB-ZF) achieved much better results than an ANalog-only BeamSTeering (AN-BST) scheme with no baseband precoder. In fact, this approach was capable of better approaching the upper bound defined by the singular value decomposition precoder (SVD-UB). Other relevant conclusion is that the design of precoding algorithms should be adapted to the communication schemes. While considering all the specific constraints may allow the maximization of the system performance of the system, formulating and solving the corresponding optimization problem may not be so simple.

Motivated by the work above, in this paper we developed an algorithm for hybrid precoding design which can accommodate different low-complexity architectures suitable for both mmWave and THz MU MIMO systems. It is based on the idea of accomplishing a near-optimal approximation of the fully digital precoder for any configuration of antennas, RF chains and data streams through the application of the alternating direction method of multipliers (ADMM) [22]. ADMM is a well-known and effective method for solving convex optimization problems but can also be a powerful heuristic for several non-convex problems [22], [23]. To use it effectively within the context of MU MIMO, proper formulation of the hybrid design problem as a multiple constrained matrix factorization problem is first presented. Using the proposed formulation, an iterative algorithm comprising several reduced complexity steps is obtained. The main contributions of this paper can be summarized as follows:

- We propose a hybrid design algorithm with near fully digital performance, where the digital precoder, analog precoder and multiuser interference mitigation are computed separately through simple closed-form solutions. The hybrid design algorithm is developed independently of a specific channel or antenna configuration, which allows its application in mmWave and THz system. Whereas our previous work [10] also proposed an hybrid design algorithm for mmWave, it did not address multiuser systems, and in particular the MIMO broadcast channel. Therefore, it does not include any step for inter-user interference mitigation within its design. As we show in here, for this multiuser channel the hybrid design method must also deal with the residual inter-user interference as it can degrade system performance, particularly at high SNRs.

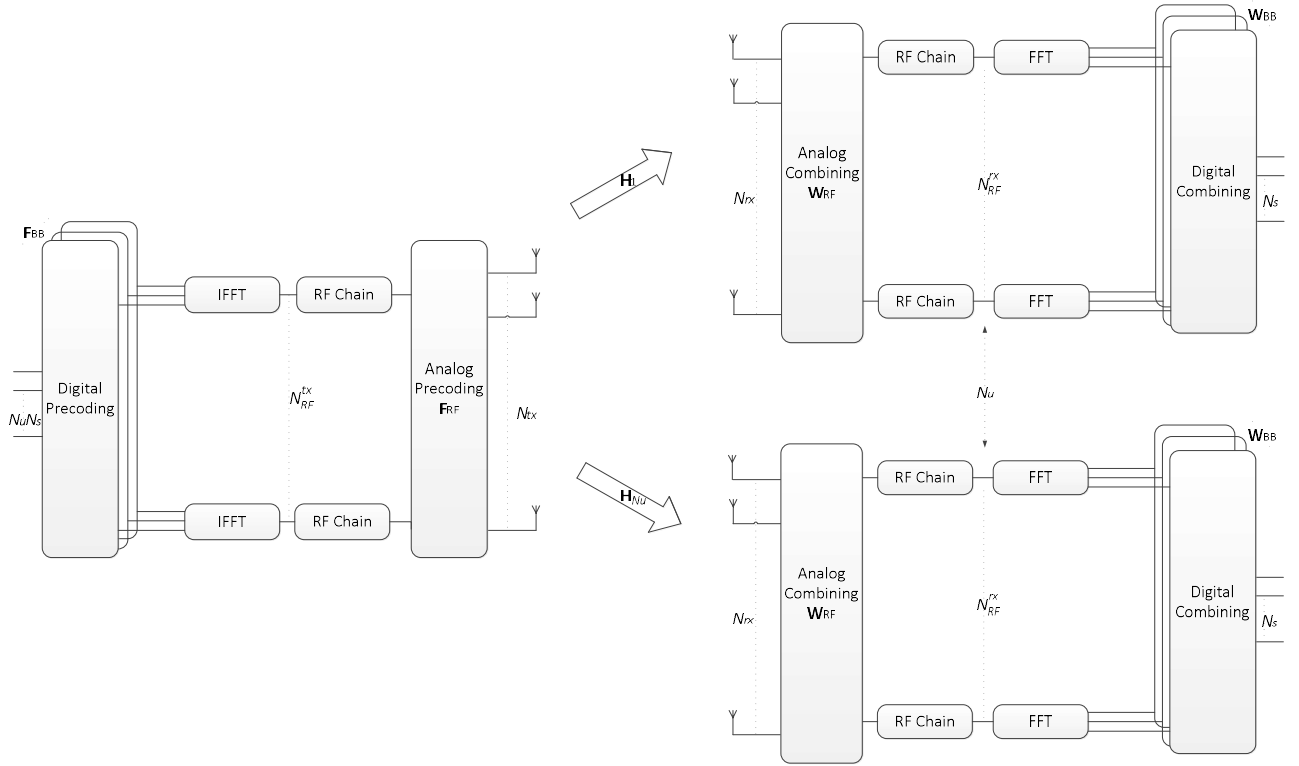

Fig. 1. A multiuser OFDM mmWave/THz MIMO system with hybrid precoding.

- Due to the separability of the different steps (analog precoder, digital precoder and interference suppression), the proposed algorithm can incorporate different architectures making it suitable for supporting UM-MIMO in severely hardware constrained THz systems. Unlike [10], where we only considered de adoption of phase-shifters, in this paper we present explicit solutions for some of the most common architectures, namely FC, AoSA and DAoSA structures based in either Unquantized Phase Shifters (UPS), Quantized Phase Shifters (QPS), Switches (Swi), Switches and Inverters (SI), Antenna Selection (AS) or Double Phase Shifters (DPS);

- To cope with the large bandwidths available in mmWave/THz bands, where practical MIMO systems likely have to operate in frequency selective channels, the proposed hybrid design considers the application in a multicarrier context, where the same analog precoder is applied at the different frequencies;

- We explicitly show how the proposed design can be applied to a DAoSAs approach, where a reduced number of switches are inserted at each AoSA panel which allows the connections to the RF chains to be dynamically adjusted. Through extensive simulations it is shown that our proposed solution is capable of achieving good trade-offs between spectral efficiency, hardware complexity and power consumption, proving to be a suitable solution for the deployment of ultra-massive MIMO especially in hardware constrained THz systems.

The paper is organized as follows: section II presents the adopted system model. The adopted formulation of the hybrid design problem for the MU MIMO scenario and the proposed algorithm are described in detail in section III, which includes the implementation of the algorithm for different analog architectures. Performance results are then presented in section IV. Finally, the conclusions are outlined in section V.

*Notation:* Matrices and vectors are denoted by uppercase and lowercase boldface letters, respectively. The superscript $(.)^H$

## II. SYSTEM MODEL

In this section, we present the system and channel models adopted for the design of the hybrid precoding algorithm. Let us consider the OFDM base system illustrated in Fig. 1. In this case we have a mmWave/THz hybrid multiuser MIMO system, where a base station (BS) is equipped with $N_{tx}$ antennas and transmits to $N_u$ users equipped with $N_{rx}$ antennas over $F$ carriers, as can be seen in Fig. 1. On each subcarrier, $N_s$ data streams are transmitted to each user which are represented as $\mathbf{s}_k = \left[\mathbf{s}_{k,1}^T ... \mathbf{s}_{k,N_u}^T\right]^T$, with $\mathbf{s}_{k,u} \in \mathbb{C}^{N_s \times 1}$. Both the precoder and combiner comprise separate digital and analog processing blocks. Since the analog precoder (combiner) is located after (before) the IFFT (FFT) blocks, it is shared between the different subcarriers, as in [25], [26]. Regarding the analog precoder and combiner, which are represented by matrices $\mathbf{F}_{\text{RF}} \in \mathbb{C}^{N_{tx} \times N_{RF}^{tx}}$ and $\mathbf{W}_{\text{RF}_u} \in \mathbb{C}^{N_{rx} \times N_{RF}^{rx}}$ with $u = 1,...,N_u$, it is assumed that $N_u N_s \leq N_{RF}^{tx} \leq N_{tx}$ and $N_s \leq N_{RF}^{rx} \leq N_{rx}$, where $N_{RF}^{tx}$ and $N_{RF}^{rx}$ are the number of RF chains at the BS and each user, respectively. The received signal model at subcarrier $k$ after the combiner can be written as

$$\mathbf{Y}_{k,u} = \sqrt{\rho_u} \mathbf{W}_{\text{BB}_{k,u}}^H \mathbf{W}_{\text{RF}_u}^H \mathbf{H}_{k,u} \mathbf{F}_{\text{RF}} \mathbf{F}_{\text{BB}_k} \mathbf{s}_k +$$
$$+ \mathbf{W}_{\text{BB}_{k,u}}^H \mathbf{W}_{\text{RF}_u}^H \mathbf{n}_{k,u}, \quad (1)$$

where $\mathbf{H}_{k,u} \in \mathbb{C}^{N_{rx} \times N_{tx}}$ is the frequency domain channel matrix (assumed to be perfectly known at the transmitter and receiver) between the base station and the $u^{\text{th}}$ receiver at subcarrier $k$. Vector $\mathbf{n}_{k,u} \in \mathbb{C}^{N_{rx} \times 1}$ contains independent zero-mean circularly symmetric Gaussian noise samples with covariance $\sigma_n^2 \mathbf{I}_{N_{rx}}$ and $\rho_u$ denotes the average received power. The digital baseband precoders and combiners are denoted by $\mathbf{F}_{\text{BB}_k} \in \mathbb{C}^{N_{RF}^{tx} \times N_u N_s}$ and $\mathbf{W}_{\text{BB}_{k,u}} \in \mathbb{C}^{N_{RF}^{rx} \times N_s}$, respectively. Regarding the channel model, it is important to note that even though the mmWave and THz bands share a few commonalities, the THz channel has several peculiarities that distinguish it from the mmWave channel. For example, the very high scattering and diffraction losses in the THz band will typically result in a much sparser channel in the angular domain with fewer multipaths components (typically less than 10) [21]. Furthermore, the gap between the line of sight (LOS) and non-line of sight (NLOS) components tends to be very large making it often LOS-dominant with NLOS-assisted [26]. An additional aspect relies on the much larger bandwidth of THz signals which can suffer performance degradation due to the so-called beam split effect, where the transmission paths squint into different spatial directions depending on the subcarrier frequency [21]. In light of this, in this paper we consider a clustered wideband geometric channel, which is commonly adopted both in mmWave [15] and THz literature [20], [26], [27], [29]. However, it should be noted that the hybrid precoding/combining approach proposed in this paper is independent of a specific MIMO channel. In this case the frequency domain channel matrices can be characterized as

$$\mathbf{H}_{k,u} = \gamma \Big( \alpha_u^{LOS} \mathbf{a}_r \big( \phi_u^{r,LOS}, \theta_u^{r,LOS} \big) \mathbf{a}_t \big( \phi_u^{t,LOS}, \theta_u^{t,LOS} \big)^H +$$
$$+ \sum_{i=1}^{N_{cl}} \sum_{l=1}^{N_{ray}} \alpha_{i,l,u} \mathbf{a}_r \big( \phi_{i,l,u}^r, \theta_{i,l,u}^r \big) \mathbf{a}_t \big( \phi_{i,l,u}^t, \theta_{i,l,u}^t \big)^H \Big) e^{-j2\pi \tau_{i,u} f_k}, \quad (2)$$

where $N_{cl}$ denotes the scattering clusters with each cluster $i$ having a time delay of $\tau_{i,u}$ and $N_{ray}$ propagations paths. $\alpha_u^{LOS}$ $\alpha_{i,l,u}$ are the complex gains of the LOS component and of the $l^{\text{th}}$ ray from cluster $i$. Index $u$ is the user ($u = 1,...,N_u$), $f_k = f_c + \frac{B}{F}\left(k - 1 - \frac{F-1}{2}\right)$ ($k = 1,...,F$) is the $k^{\text{th}}$ subcarrier frequency, $B$ is the bandwidth, $f_c$ is the central frequency and $\gamma$ is a normalizing factor such that $\mathrm{E}\Big[\|\mathbf{H}_{k,u}\|_F^2\Big] = N_{tx} N_{rx}$. Vectors $\mathbf{a}_t\big(\phi_{i,l,u}^t, \theta_{i,l,u}^t\big)$ and $\mathbf{a}_r\big(\phi_{i,l,u}^r, \theta_{i,l,u}^r\big)$ represent the transmit and receive antenna array responses at the azimuth and elevation angles of $\big(\phi_{i,l,u}^t, \theta_{i,l,u}^t\big)$ and $\big(\phi_{i,l,u}^r, \theta_{i,l,u}^r\big)$, respectively. Vectors $\mathbf{a}_t\big(\phi_u^{t,LOS}, \theta_u^{t,LOS}\big)$ and $\mathbf{a}_r\big(\phi_u^{r,LOS}, \theta_u^{r,LOS}\big)$ have similar meanings but refer to the LOS path angles $\big(\phi_u^{t,LOS}, \theta_u^{t,LOS}\big)$ and $\big(\phi_u^{r,LOS}, \theta_u^{r,LOS}\big)$. By carefully selecting the parameters of the channel model we can make it depict a mmWave or a THz channel. Considering Gaussian signaling, the spectral efficiency achieved by the system for the transmission to MS-$u$ in subcarrier $k$ is [29]

$$\mathbf{R}_{k,u} = \log_2 \Big| \mathbf{I}_{N_{RF}^{rx}} + \mathbf{R}_u^{-1} \mathbf{W}_{\text{BB}_{k,u}}^H \mathbf{W}_{\text{RF}_u}^H \mathbf{H}_{k,u} \mathbf{F}_{\text{RF}} \mathbf{F}_{\text{BB}_{k,u}} \times$$
$$\times \mathbf{F}_{\text{BB}_{k,u}}^H \mathbf{F}_{\text{RF}}^H \mathbf{H}_{k,u}^H \mathbf{W}_{\text{RF}_u} \mathbf{W}_{\text{BB}_{k,u}} \Big| \quad (3)$$

where $\mathbf{R}_{k,u}$ is the covariance matrix of the total inter-user interference plus noise at MS-$u$, which is characterized by

$$\mathbf{R}_{k,u} = \mathbf{W}_{\text{BB}_{k,u}}^H \mathbf{W}_{\text{RF}_u}^H (\mathbf{H}_{k,u} \sum_{j \neq u}^{N_u} \mathbf{F}_{\text{RF}} \mathbf{F}_{\text{BB}_{k,j}} \mathbf{F}_{\text{BB}_{k,j}}^H \mathbf{F}_{\text{RF}}^H \mathbf{H}_{k,u}^H +$$
$$+ \sigma^2 \mathbf{I}_{N_{rx}}) \mathbf{W}_{\text{RF}_u} \mathbf{W}_{\text{BB}_{k,u}}, \quad (4)$$

III. PROPOSED HYBRID DESIGN ALGORITHM

In this section, we will introduce the algorithm for the hybrid precoding problem and show how it can be adapted to different architectures. Although we will focus on the precoder design, a similar approach can be adopted for the combiner. However, since our design assumes that inter-user interference suppression is applied at the transmitter, only single-user detection is required at the receiver and therefore the algorithm reduces to the one described in [10].

A. Main Algorithm

Although there are several problem formulations for the hybrid design proposed in the literature, one of the most effective relies on the minimization of the Frobenius norm of the difference between the fully digital precoder and the hybrid precoder [22], [30], [31], [32]. In this paper we follow this matrix approximation-based approach which can be formulated as

$$\min_{\mathbf{F}_{RF},\mathbf{F}_{BB_k}} \sum_{k=1}^{F} \left\| \mathbf{F}_{opt_k} - \mathbf{F}_{RF}\mathbf{F}_{BB_k} \right\|_F^2 \quad (5)$$

$$\text{subject to } \mathbf{F}_{RF} \in \mathcal{C}_{N_{tx} \times N_{RF}^{tx}} \quad (6)$$

$$\left\| \mathbf{F}_{RF}\mathbf{F}_{BB_k} \right\|_F^2 \leq N_u N_s \quad (7)$$

where $\mathcal{C}_{N_{tx} \times N_{RF}}$ is the set of feasible analog precoding matrices, which is defined according to the adopted RF architecture (it will be formally defined for several different architectures in the next subsection). Matrix $\mathbf{F}_{opt_k}$ denotes the fully digital precoder which can be designed so as to enforce zero inter-user interference using for example the block-diagonalization approach described in in [33]. Even if $\mathbf{F}_{opt_k}$ is selected in order to cancel all interference between users, the hybrid design resulting as a solution of (5)-(7) will correspond to an approximation and, as such, residual inter-user interference will remain. To avoid the performance degradation that will result from this, an additional constraint can be added to the problem formulation, namely

$$\sum_{\substack{u'=1 \\ u' \neq u}}^{N_u} \mathbf{H}_{k,u'} \mathbf{F}_{RF} \mathbf{F}_{BB_{k,u}} = 0, \quad k=1,...,F, \quad u=1,...,N_u \quad (8)$$

where $\mathbf{F}_{BB_{k,u}} = \mathbf{F}_{BB_k}[:,(u-1)N_s+1:uN_s]$. This restriction is equivalent to enforcing $\mathbf{F}_{RF}\mathbf{F}_{BB_{k,u}}$ to lie in the null space of $\bar{\mathbf{H}}_{k,u} \in \mathbb{C}^{(N_u-1)N_{rx} \times N_{tx}}$ ($\bar{\mathbf{H}}_{k,u}$ is a matrix corresponding to $\mathbf{H}_k$ with the $N_{rx}$ lines of user $u$ removed) which we denote as $\mathcal{N}(\bar{\mathbf{H}}_{k,u})$. The overall optimization problem can be then expressed as

$$\min_{\mathbf{F}_{RF},\mathbf{F}_{BB_k}} \sum_{k=1}^{F} \left\| \mathbf{F}_{opt_k} - \mathbf{F}_{RF}\mathbf{F}_{BB_k} \right\|_F^2 \quad (9)$$

$$\text{subject to } \mathbf{F}_{RF} \in \mathcal{C}_{N_{tx} \times N_{RF}^{tx}} \quad (10)$$

$$\left\| \mathbf{F}_{RF}\mathbf{F}_{BB_k} \right\|_F^2 \leq N_u N_s \quad (11)$$

$$\mathbf{F}_{RF}\mathbf{F}_{BB_{k,u}} \in \mathcal{N}(\bar{\mathbf{H}}_{k,u}), \quad k=1,...,F, \quad u=1,...,N_u \quad (12)$$

To derive a hybrid precoder/design algorithm that can cope with the different RF architectures we can integrate the RF constraint directly into the objective function of the optimization problem. This can be accomplished through the addition of an auxiliary variable, $\mathbf{R}$, combined with the use of the indicator function. The indicator function for a generic set $\mathcal{A}$ is defined as $I_\mathcal{A}(\mathbf{x})$, returning 0 if $\mathbf{x} \in \mathcal{A}$ and $+\infty$ otherwise. A similar approach can be adopted for integrating the other constraints, (11) and (12), also into the objective function. The optimization problem can then be rewritten as

$$\min_{\mathbf{F}_{RF},\mathbf{F}_{BB_k},\mathbf{R},\mathbf{B}_k,\mathbf{F}_{aprox_{k,u}}} \sum_{k=1}^{F} \left\| \mathbf{F}_{opt_k} - \mathbf{F}_{RF}\mathbf{F}_{BB_k} \right\|_F^2 + I_\mathcal{C}(\mathbf{R}) + $$
$$+ \sum_{k=1}^{F} I_{\|\cdot\|_F^2 = N_u N_s}(\mathbf{B}_k) + \sum_{k=1}^{F}\sum_{u=1}^{N_u} I_{\mathcal{N}(\bar{\mathbf{H}}_{k,u})}(\mathbf{F}_{aprox_{k,u}}) \quad (13)$$

$$\text{subject to } \mathbf{R} = \mathbf{F}_{RF} \quad (14)$$

$$\mathbf{B}_k = \mathbf{F}_{RF}\mathbf{F}_{BB_k} \quad (15)$$

$$\mathbf{F}_{aprox_{k,u}} = \mathbf{F}_{RF}\mathbf{F}_{BB_k} \quad (16)$$

where $\mathbf{F}_{aprox_k} = \left[ \mathbf{F}_{aprox_{k,1}}, \cdots, \mathbf{F}_{aprox_{k,N_u}} \right]$. The augmented Lagrangian function (ALF) for (13)-(16) can be written as

$$L_{\rho,\eta,\mu}(\mathbf{F}_{RF},\mathbf{F}_{BB},\mathbf{R},\mathbf{B},\mathbf{F}_{aprox},\mathbf{\Lambda},\mathbf{\Psi},\mathbf{\Gamma}) = \sum_{k=1}^{F} \left\| \mathbf{F}_{opt_k} - \mathbf{F}_{RF}\mathbf{F}_{BB_k} \right\|_F^2 + $$
$$+ I_\mathcal{C}(\mathbf{R}) + \sum_{k=1}^{K} I_{\|\cdot\|_F^2 = N_u N_s}(\mathbf{B}_k) + \sum_{k=1}^{F}\sum_{u=1}^{Nu} I_{\mathcal{N}(\bar{\mathbf{H}}_{k,u})}(\mathbf{F}_{aprox_{k,u}}) + $$
$$+ 2\operatorname{Re}\{\operatorname{tr}(\mathbf{\Lambda}^H(\mathbf{F}_{RF}-\mathbf{R}) + \operatorname{tr}(\mathbf{\Psi}_k^H \sum_{k=1}^{F}(-\mathbf{B}_k + \mathbf{F}_{RF}\mathbf{F}_{BB_k}))$$
$$+ \operatorname{tr}(\mathbf{\Gamma}_k^H \sum_{k=1}^{F}(-\mathbf{F}_{aprox_k} + \mathbf{F}_{RF}\mathbf{F}_{BB_k}))\} + \rho \left\| \mathbf{F}_{RF}-\mathbf{R} \right\|_F^2 + $$
$$+ \eta \sum_{k=1}^{F} \left\| -\mathbf{B}_k + \mathbf{F}_{RF}\mathbf{F}_{BB_k} \right\|_F^2 + \mu \sum_{k=1}^{F} \left\| -\mathbf{F}_{aprox_k} + \mathbf{F}_{RF}\mathbf{F}_{BB_k} \right\|_F^2, \quad (17)$$

where $\mathbf{\Lambda} \in \mathbb{C}^{N_{tx} \times N_{RF}}$, $\mathbf{\Psi} \in \mathbb{C}^{N_{tx} \times N_{RF}}$ and $\mathbf{\Gamma} \in \mathbb{C}^{N_{tx} \times N_{RF}}$ are dual variables and $\rho$, $\eta$, $\mu$ are penalty parameters. After some straightforward algebraic manipulation and working with scaled dual variables $\mathbf{U} = 1/\rho \cdot \mathbf{\Lambda}$, $\mathbf{W}_k = 1/\eta \cdot \mathbf{\Psi}$ and $\mathbf{Z}_k = 1/\mu \cdot \mathbf{\Gamma}_k$ we can rewrite the ALF as

$$L_{\rho,\eta,\mu}(\mathbf{F}_{RF},\mathbf{F}_{BB},\mathbf{R},\mathbf{B},\mathbf{F}_{aprox},\mathbf{U},\mathbf{W},\mathbf{Z}) = \sum_{k=1}^{K} \left\| \mathbf{F}_{opt_k} - \mathbf{F}_{RF}\mathbf{F}_{BB_k} \right\|_F^2 + $$
$$+ I_\mathcal{C}(\mathbf{R}) + \sum_{k=1}^{F} I_{\|\cdot\|_F^2 = N_u N_s}(\mathbf{B}_k) + \sum_{k=1}^{F}\sum_{u=1}^{Nu} I_{\mathcal{N}(\bar{\mathbf{H}}_{k,u})}(\mathbf{F}_{aprox_{k,u}}) + $$
$$+ \rho \left\| \mathbf{F}_{RF}-\mathbf{R}+\mathbf{U} \right\|_F^2 - \rho \left\| \mathbf{U} \right\|_F^2 + $$
$$+ \eta \sum_{k=1}^{F} \left\| -\mathbf{B}_k + \mathbf{F}_{RF}\mathbf{F}_{BB_k} + \mathbf{W}_k \right\|_F^2 - \eta \sum_{k=1}^{F} \left\| \mathbf{W}_k \right\|_F^2 + $$
$$+ \mu \sum_{k=1}^{F} \left\| -\mathbf{F}_{aprox_k} + \mathbf{F}_{RF}\mathbf{F}_{BB_k} + \mathbf{Z}_k \right\|_F^2 - \mu \sum_{k=1}^{F} \left\| \mathbf{Z}_k \right\|_F^2, \quad (18)$$

Based on the ADMM [22], we can apply the gradient ascent to the dual problem involving the ALF, which allows us to obtain an iterative precoding algorithm comprising the following sequence of steps. We start with the minimization of the ALF over $\mathbf{F}_{RF}$ for iteration $t+1$ defined as

$$\mathbf{F}_{RF}^{(t+1)} = \min_{\mathbf{F}_{RF}} L_{\rho,\eta,\mu}(\mathbf{F}_{RF},\mathbf{F}_{BB}^{(t)},\mathbf{R}^{(t)},\mathbf{B}^{(t)},....$$
$$...,\mathbf{F}_{aprox}^{(t)},\mathbf{U}^{(t)},\mathbf{W}^{(t)},\mathbf{Z}^{(t)}). \quad (19)$$

which can be obtained from

$$\nabla_{\mathbf{F}_{\text{RF}}^H} L_{\rho,\eta,\mu}(\mathbf{F}_{\text{RF}},\mathbf{F}_{\text{BB}_k}^{(t)},\mathbf{R}^{(t)},\mathbf{B}^{(t)},....$$
$$....,\mathbf{F}_{\text{aprox}}^{(t)},\mathbf{U}^{(t)},\mathbf{W}^{(t)},\mathbf{Z}^{(t)})=0 \quad (20)$$

leading to the closed form expression

$$\mathbf{F}_{\text{RF}}^{(t+1)} = \left[\sum_{k=1}^{F}[\mathbf{F}_{\text{opt}_k} + \eta(\mathbf{B}_k^{(t)} - \mathbf{W}_k^{(t)}) + \mu(\mathbf{F}_{\text{aprox}}^{(t)} - \mathbf{Z}_k^{(t)})] \times \right.$$
$$\times \mathbf{F}_{\text{BB}_k}^{(t)\,H} + \rho\left(\mathbf{R}^{(t)} - \mathbf{U}^{(t)}\right)\right] \times$$
$$\times [(1+\eta+\mu)\sum_{k=1}^{F} \mathbf{F}_{\text{BB}_k}^{(t)} \mathbf{F}_{\text{BB}_k}^{(t)\,H} + \rho \mathbf{I}_{N_{\text{RF}}^{tx}}]^{-1}. \quad (21)$$

After obtaining the expression for $\mathbf{F}_{\text{RF}}$, $\mathbf{F}_{\text{BB}}^{(t+1)}$ can be found by following the same methodology. In this case the minimization is expressed as

$$\mathbf{F}_{\text{BB}}^{(t+1)} = \min_{\mathbf{F}_{\text{BB}}} L_{\rho,\eta,\mu}(\mathbf{F}_{\text{RF}}^{(t)},\mathbf{F}_{\text{BB}},\mathbf{R}^{(t)},\mathbf{B}^{(t)},....$$
$$....,\mathbf{F}_{\text{aprox}}^{(t)},\mathbf{U}^{(t)},\mathbf{W}^{(t)},\mathbf{Z}^{(t)}). \quad (22)$$

from which by applying

$$\nabla_{\mathbf{F}_{\text{BB}}^H} L_{\rho,\eta,\mu}(\mathbf{F}_{\text{RF}}^{(t)},\mathbf{F}_{\text{BB}},\mathbf{R}^{(t)},\mathbf{B}^{(t)},....$$
$$....,\mathbf{F}_{\text{aprox}}^{(t)},\mathbf{U}^{(t)},\mathbf{W}^{(t)},\mathbf{Z}^{(t)})=0 \quad (23)$$

leads to the closed form expression

$$\mathbf{F}_{\text{BB}_k}^{(t+1)} = (1+\eta+\mu)\left(\mathbf{F}_{\text{RF}}^{(t+1)H}\mathbf{F}_{\text{RF}}^{(t+1)}\right)^{-1}\mathbf{F}_{\text{RF}}^{(t+1)H}\cdot$$
$$\cdot\left(\mathbf{F}_{\text{opt}_k} + \eta(\mathbf{B}_k^{(t)} - \mathbf{W}_k^{(t)}) + \mu(\mathbf{F}_{\text{aprox}_k}^{(t)} - \mathbf{Z}_k^{(t)})\right), k=1,...,F \quad (24)$$

The next steps consist of the minimization over $\mathbf{R}$ and $\mathbf{B}_k$. The minimization of (18) with respect to $\mathbf{R}$ and $\mathbf{B}_k$ can be written as

$$\mathbf{R}^{(t+1)} = \min_{\mathbf{R}}\left\{\mathrm{I}_{\mathcal{C}_{N_{tx} \times N_{RF}}}(\mathbf{R}) + \rho\left\|\mathbf{F}_{RF}^{(t+1)} - \mathbf{R} + \mathbf{U}^{(t)}\right\|_F^2\right\}$$
$$= \Pi_{\mathcal{C}_{N_{tx} \times N_{RF}}}\left(\mathbf{F}_{RF}^{(t+1)} + \mathbf{U}^{(t)}\right) \quad (25)$$

and

$$\mathbf{B}_k^{(t+1)} = \min_{\mathbf{R}}\left\{\mathrm{I}_{\|\cdot\|_F^2 = N_u N_s}(\mathbf{B}_k) + \eta\left\|\mathbf{F}_{RF}^{(t+1)}\mathbf{F}_{BB_k}^{(t+1)} - \mathbf{B}_k + \mathbf{W}^{(t)}\right\|_F^2\right\}$$
$$= \Pi_{\|\cdot\|_F^2 = N_u N_s}\left(\mathbf{F}_{RF}^{(t+1)}\mathbf{F}_{BB_k}^{(t+1)} + \mathbf{W}^{(t)}\right), \quad k=1,...,F. \quad (26)$$

where $\Pi_{\mathcal{C}_{a \times b}}(\cdot)$ and $\Pi_{\|\cdot\|_F^2 = N_u N_s}(\cdot)$ denote the projection onto set $\mathcal{C}_{a \times b}$ and onto the set of matrices whose squared Frobenius norm is $N_u N_s$, respectively. While the former projection depends on the adopted analog architecture and will be explained in the next subsection, the second projection is simply computed as

$$\mathbf{B}_k^{(t+1)} = \frac{\left(\mathbf{F}_{RF}^{(t+1)}\mathbf{F}_{BB_k}^{(t+1)} + \mathbf{W}^{(t)}\right)\sqrt{N_u N_s}}{\left\|\mathbf{F}_{RF}^{(t+1)}\mathbf{F}_{BB_k}^{(t+1)} + \mathbf{W}^{(t)}\right\|_F} \quad (27)$$

The minimization of (18) with respect of $\mathbf{F}_{\text{aprox}_{k,u}}$ can be written as

$$\mathbf{F}_{\text{aprox}_{k,u}}^{(t+1)} = \min_{\mathbf{F}_{\text{aprox}_{k,u}}}\{\mathrm{I}_{\mathcal{N}(\bar{\mathbf{H}}_{k,u})}(\mathbf{F}_{\text{aprox}_{k,u}}) +$$
$$+\mu\left\|\mathbf{F}_{RF}^{(t+1)}\mathbf{F}_{BB_{k,(u-1)N_s+1:uN_s}}^{(t+1)} - \mathbf{F}_{\text{aprox}_{k,u}} + \mathbf{Z}_{k,(u-1)N_s+1:uN_s}^{(t)}\right\|_F^2\}$$
$$= \Pi_{\mathcal{N}(\bar{\mathbf{H}}_{k,u})}\left(\mathbf{F}_{RF}^{(t+1)}\mathbf{F}_{BB_{k,(u-1)N_s+1:uN_s}}^{(t+1)} + \mathbf{Z}_{k,(u-1)N_s+1:uN_s}^{(t)}\right) \quad (28)$$

which also involves a projection, $\Pi_{\mathcal{N}(\bar{\mathbf{H}}_{k,u})}(\cdot)$, but in this case onto the null-space of $\bar{\mathbf{H}}_{k,u}$. Let us use $\mathbf{A}$ to denote $\mathbf{A} = \mathbf{F}_{RF}^{(t+1)}\mathbf{F}_{BB_{k,u}}^{(t+1)} + \mathbf{Z}_{k,(u-1)N_s+1:uN_s}^{(t)}$. The procedure to compute the projection of matrix $\mathbf{A}$ onto the null-space of $\bar{\mathbf{H}}_{k,u}$ can be formulated as another optimization problem, which can be expressed as

$$\min \sum_{i=1}^{N_s}\left\|\mathbf{A}_{:,i} - \mathbf{X}_{:,i}\right\|_F^2 \quad (29)$$
$$\text{subject to } \bar{\mathbf{H}}_{k,u}\mathbf{X}_{:,i} = 0 \quad (30)$$

The general solution for this problem is presented in [30] corresponding to

$$\mathbf{X}_{:,i} = \left(\mathbf{I}_{N_{tx}} - \bar{\mathbf{H}}_{k,u}^H\left(\bar{\mathbf{H}}_{k,u}\bar{\mathbf{H}}_{k,u}^H\right)^{-1}\bar{\mathbf{H}}_{k,u}\right)\mathbf{A}_{:,i}, \quad i=1,...,N_s \quad (31)$$

Reordering the column vectors in the original matrix form results the final expression which can be rewritten as

$$\mathbf{X} = \left(\mathbf{I}_{N_{tx}} - \bar{\mathbf{H}}_{k,u}^H\left(\bar{\mathbf{H}}_{k,u}\bar{\mathbf{H}}_{k,u}^H\right)^{-1}\bar{\mathbf{H}}_{k,u}\right)\mathbf{A}$$
$$= \left(\mathbf{I}_{N_{tx}} - \bar{\mathbf{V}}_{k,u}^{(1)}\left(\bar{\mathbf{V}}_{k,u}^{(1)}\right)^H\right)\mathbf{A} \quad (32)$$

In this expression, $\bar{\mathbf{V}}_{k,u}^{(1)}$ denotes the matrix containing the right singular vectors corresponding to the nonzero singular values associated to the singular value decomposition (SVD) given by $\bar{\mathbf{H}}_{k,u} = \bar{\mathbf{U}}_{k,u}\bar{\mathbf{\Lambda}}_{k,u}\left[\bar{\mathbf{V}}_{k,u}^{(1)} \ \bar{\mathbf{V}}_{k,u}^{(0)}\right]^H$.

Therefore, to compute matrix $\mathbf{X}$ one can perform a single value decomposition of $\bar{\mathbf{H}}_{k,u}$ and then use this to remove the projection of $\mathbf{A}$ onto the row space of $\bar{\mathbf{H}}_{k,u}$. Finally, the expressions for the update of dual variables $\mathbf{U}$, $\mathbf{W}$ and $\mathbf{Z}$ are given by

$$\mathbf{U}^{(t+1)} = \mathbf{U}^{(t)} + \mathbf{F}_{RF}^{(t+1)} - \mathbf{R}^{(t+1)} \quad (33)$$

$$\mathbf{W}_k^{(t+1)} = \mathbf{W}_k^{(t)} + \mathbf{F}_{RF}^{(t+1)}\mathbf{F}_{BB_k}^{(t+1)} - \mathbf{B}_k^{(t+1)} \quad (34)$$

$$\mathbf{Z}_k^{(t+1)} = \mathbf{Z}_k^{(t)} + \mathbf{F}_{\text{RF}}^{(t+1)} \mathbf{F}_{\text{BB}_k}^{(t+1)} - \mathbf{F}_{\text{aprox}_k}^{(t+1)} \qquad (35)$$

Appropriate values for the penalty parameters can be obtained in a heuristic manner by performing numerical simulations. Regarding the initialization and termination of the algorithm, the same approach described in [10] can be adopted. The whole algorithm is summarized in Table I. In this table, $Q$ denotes the maximum number of iterations.

TABLE I
GENERAL ITERATIVE HYBRID DESIGN ALGORITHM.

1: **Input:** $\mathbf{F}_{\text{opt}_k}$, $\mathbf{F}_{\text{RF}}^{(0)}$, $\mathbf{F}_{\text{BB}_k}^{(0)}$, $\mathbf{R}^{(0)}$, $\mathbf{B}_k^{(0)}$, $\mathbf{F}_{\text{aprox}_{k,u}}^{(0)}$, $\rho$, $Q$
2: **for** $t=0, 1, \ldots, Q-1$ **do**
3: Compute $\mathbf{F}_{\text{RF}}^{(t+1)}$ using (21).
4: Compute $\mathbf{F}_{\text{BB}_k}^{(t+1)}$ using (24), for all $k=1, \ldots, F$.
5: Compute $\mathbf{R}^{(t+1)}$ using (25).
6: Compute $\mathbf{B}_k^{(t+1)}$ using (26), for all $k=1, \ldots, F$.
7: Compute $\mathbf{F}_{\text{aprox}_{k,u}}^{(t+1)}$ using (28), for all $k=1, \ldots, F$ and $u=1, \ldots, N_u$.
8: Update $\mathbf{U}^{(t+1)}$ using (33).
9: Update $\mathbf{W}_k^{(t+1)}$ using (34), for all $k=1, \ldots, F$.
10: Update $\mathbf{Z}_k^{(t+1)}$ using (35), for all $k=1, \ldots, F$.
11: **end for.**
12: $\hat{\mathbf{F}}_{\text{RF}} \leftarrow \mathbf{R}^{(Q)}$.
13: $\hat{\mathbf{F}}_{\text{BB}_k} \leftarrow \left(\hat{\mathbf{F}}_{\text{RF}}^H \hat{\mathbf{F}}_{\text{RF}}\right)^{-1} \hat{\mathbf{F}}_{\text{RF}}^H \mathbf{F}_{\text{aprox}_k}^{(Q)}$, for all $k=1, \ldots, F$.
14: $\hat{\mathbf{F}}_{\text{BB}_k} \leftarrow \sqrt{N_u N_s} \left\|\hat{\mathbf{F}}_{\text{BB}_k}^H \hat{\mathbf{F}}_{\text{BB}_k}\right\|_F^{-1} \hat{\mathbf{F}}_{\text{BB}_k}$.
15: **Output:** $\hat{\mathbf{F}}_{\text{RF}}$, $\hat{\mathbf{F}}_{\text{BB}}$.

The projection operation is the only step specific to the implemented architecture, as will be explained in the next subsection. The projection operation is the only step specific to the implemented architecture, as will be explained in the next subsection.

### B. Analog Rf Precoder/Combiner Structure

The projection required for obtaining matrix R in step 5 of the precoding algorithm, has to be implemented according to the specific analog beamformer [6], [20], [34]-[38]. This makes the proposed scheme very generic, allowing it to be easily adapted to different RF architectures. In the following we will consider a broad range of architectures that can be adopted at the RF precoder for achieving reduced complexity and power consumption implementations. We will consider FC, AoSA and DAoSA structures as illustrated in Fig. 2. Besides phase shifters, we will also consider several alternative implementations for these structures, as shown in Fig. 3.

*1) Unquantized Phase Shifters (UPS)*

In the first case we consider the use of infinite resolution phase shifter. For this architecture the RF constraint set is given by

$$\mathcal{C}_{a \times b} = \left\{\mathbf{X} \in \mathbb{C}^{a \times b} : |X_{i,j}| = 1\right\} \qquad (36)$$

and the corresponding projection can be performed simply using

$$\mathbf{R}^{(t+1)} = \left(\mathbf{F}_{\text{RF}}^{(t+1)} + \mathbf{W}^{(t)}\right) \oslash \left|\mathbf{F}_{\text{RF}}^{(t+1)} + \mathbf{W}^{(t)}\right|. \qquad (37)$$

*2) Quantized Phase Shifters (QPS)*

The second case considers a more realistic scenario, in which phase shifters can be digitally controlled with $N_b$ bits. These devices allow the selection of $2^{N_b}$ different quantized phases and the RF constraint set becomes

$$\mathcal{C}_{a \times b} = \left\{\mathbf{X} \in \mathbb{C}^{a \times b} : X_{i,j} = e^{2\pi k i / 2^{N_b}}, k=0,\ldots,2^{N_b}-1\right\} \qquad (38)$$

The implementation of the projection in line 5 of Table I can be obtained as the following element-wise quantization

$$R_{i,j}^{(t+1)} = e^{\min_{k=0,\ldots,2^{N_b}-1}\left\{\text{angle}\left(F_{\text{RF}_{i,j}}^{(t+1)} + W_{i,j}^{(t)}\right) - 2\pi k / 2^{N_b}\right\}},$$

$$i = 1,\ldots,N_{tx}, j = 1,\ldots,N_{RF}^{tx}, \qquad (39)$$

*3) Switches and inverters (SI)*

Assuming that $N_b = 1$, then each variable phase shifter of the previous architecture can be replaced by a pair of switched lines, including also an inverter. The corresponding constraint set can be reduced to

$$\mathcal{C}_{a \times b} = \left\{\mathbf{X} \in \mathbb{R}^{a \times b} : X_{i,j} = \pm 1\right\} \qquad (40)$$

and the implementation of the projection simplifies to

$$R_{i,j}^{(t+1)} = \text{sign}\left(\text{Re}\left[F_{\text{RF}_{i,j}}^{(t+1)} + W_{i,j}^{(t)}\right]\right) \qquad (41)$$

*4) Switches (Swi)*

Alternatively, each of the variable phase shifters can be replaced by a switch. This simplification results in a network of switches connecting each RF chain to the antennas. The RF constraint set can be represented as

$$\mathcal{C}_{a \times b} = \left\{\mathbf{X} \in \mathbb{R}^{a \times b} : X_{i,j} = 0 \lor X_{i,j} = 1\right\} \qquad (42)$$

and the projection can be implemented elementwise as

$$R_{i,j}^{(t+1)} = 1/2 + 1/2 \cdot \text{sign}\left(2\text{Re}\left[F_{\text{RF}_{i,j}}^{(t+1)} + W_{i,j}^{(t)}\right] - 1\right) \qquad (43)$$

*5) Antenna Selection (AS)*

The simplest scenario that we can consider corresponds to an architecture, where each RF chain can be only connected to a single antenna (and vice-versa). The RF constraint set will comprise a matrix with only one nonzero element per column and per row, i.e.,

$$\mathcal{C}_{a \times b} = \left\{\mathbf{X} \in \mathbb{R}^{a \times b} : X_{i,j} = 0 \lor X_{i,j} = 1, \|X_{i,:}\|_0 = 1, \|X_{:,j}\|_0 = 1\right\} \qquad (44)$$

In this definition $\|.\|_0$ represents the cardinality of a vector. Defining $\mathbf{X} = \mathbf{F}_{\text{RF}}^{(t+1)} + \mathbf{W}^{(t)}$, the projection can be approximately implemented by setting all the elements in $\mathbf{X}$ as 0 except for $X_{t_j,j} = 1$, where $t_j$ is the row position with the highest real component in column $j$:

$$t_j = \arg\max_{i=1,\dots,N_{tx}}\{\text{Real}[X_{i,j}]\} \quad (45)$$

The computation of $t_j$ is performed for all columns $j=1, \dots, N_{RF}^{tx}$, sorted by descending order in terms of highest real components. It should be noted that during this operation, the same row cannot be repeated.

*6) Array-of-Subarrays (AoSA)*

Within the context of UM-MIMO, one of the most appealing architectures for keeping the complexity acceptable relies on the use of AoSA, where each RF chain is only connected to one or more subsets of antennas (subarrays). Denoting the number of subarrays as $n_{SA}$, which is typically set as $n_{SA} = N_{RF}$, and the size of each subarray as $N_{tx}^{SA}$, then we have $N_{tx}^{SA} = \frac{N_{tx}}{n_{SA}} = \frac{N_{tx}}{N_{RF}}$. To limit the complexity of the architecture, each RF chain can connect to a maximum of $L_{\max}$ consecutive subarrays. In this case, the RF constraint set comprises matrices where each column has a maximum of $L_{\max}$ blocks of $N_{tx}^{SA}$ constant modulus elements, with all the remaining elements being zero. Defining $\mathbf{X} = \mathbf{F}_{\text{RF}}^{(t+1)} + \mathbf{W}^{(t)}$, the projection can be implemented by setting all the elements in $\mathbf{X}$ as 0 except for the subblocks in each column $j$ which fulfill

$$\left\|\mathbf{X}_{\{[(j-1)N_{tx}^{SA}+(i-1)N_{tx}^{SA}+1:(j-1)N_{tx}^{SA}+i\cdot N_{tx}^{SA}]-1\}\bmod N_{tx}\}+1,j}\right\|_1 > \frac{N_{tx}^{SA}}{2} \quad (46)$$

with $i = 1, \dots, L_{\max}$ and $j = 1, \dots, N_{RF}$. In this case, the corresponding elements of $\mathbf{R}$ are set as $R_{i,j}^{(t+1)} = (X_{i,j})/|X_{i,j}|$, assuming UPS in these connections. Clearly, the phase shifters can be replaced by any of the other alternatives presented previously.

*7) Dynamic Array-of-Subarrays (DAoSA)*

As a variation of the previous AoSA architecture, we also consider an implementation where each subarray can be connected to a maximum of $L_{\max}$ RF chains (which can be non-adjacent). In this case, the constraint set comprises matrices where each $N_{tx}^{SA} \times N_{RF}$ component submatrix contains a maximum of $L_{\max}$ columns with constant modulus elements. The rest of the matrix contains only zeros. In this case, starting with X=0, the projection can be obtained by selecting the $L_{\max}$ columns of

$$\left\|\mathbf{X}_{\{[(j-1)N_{tx}^{SA}+(i-1)N_{tx}^{SA}+1:(j-1)N_{tx}^{SA}+i\cdot N_{tx}^{SA}]-1\}\bmod N_{tx}\}+1,j}\right\|_1 > \frac{N_{tx}^{SA}}{2} \quad (47)$$

where $j = 1, \dots, n_{SA}$ with the largest $\ell_1$-norm and setting the corresponding elements of R as $R_{i,j}^{(t+1)} = (X_{i,j})/|X_{i,j}|$, assuming the use of UPS. Care must be taken to guarantee that at least one subblock will be active in every column of R. Similarly to the AoSA, the phase shifters can be replaced by any of the other presented alternatives.

*8) Double Phase Shifters (DPS)*

Another appealing architecture relies on the use of double phase shifters (DPS) since these remove the constant modulus restriction on the elements of $\mathbf{F}_{\text{RF}}$, following the idea in [38]. In this case the projection can be implemented elementwise simply as

$$R_{i,j}^{(t+1)} = \left(F_{\text{RF}_{i,j}}^{(t+1)} + W_{i,j}^{(t)}\right) - e^{i\cdot\text{angle}\left(F_{\text{RF}_{i,j}}^{(t+1)} + W_{i,j}^{(t)}\right)} \times$$
$$\times \max\left(0, \left|F_{\text{RF}_{i,j}}^{(t+1)} + W_{i,j}^{(t)}\right| - 2\right) \quad (48)$$

Similarly to other architectures, DPS can be used not only in the fully connected approach but also in the AoSA and DAoSA cases, replacing the constant modulus setting operation.

*C. Complexity*

In the proposed algorithm, the $\mathbf{F}_{\text{RF}}^{(t+1)}$ and $\mathbf{F}_{\text{BB}}^{(t+1)}$ updates (steps 3 and 4 in Table I) are defined using closed-form expressions that encompass several matrix multiplications, sums and an $N_{RF} \times N_{RF}$ matrix inverse (with an assumed complexity order of $\mathcal{O}(N_{RF}^3)$). These steps require a complexity order of $\mathcal{O}(QN_u N_s N_{RF} N_{tx} + F^{-1}QN_{RF}^2 N_{tx})$ and $\mathcal{O}(QN_u N_s N_{RF} N_{tx} + QN_{RF}^2 N_{tx})$, respectively. The $\mathbf{R}^{(t+1)}$ update (step 5) involves simple elementwise division (assuming UPS) with $\mathcal{O}(QN_{RF} N_{tx})$ while variable $\mathbf{B}_k^{(t+1)}$ (step 6) comprises a Frobenius norm computation with $\mathcal{O}(QN_u N_s N_{RF} N_{tx})$. Step 7, the $\mathbf{F}_{\text{aprox}_{k,u}}^{(t+1)}$ update, has a complexity order of $\mathcal{O}(QN_{tx}^2 N_u N_s + N_u^3 N_{tx} N_{rx}^2 + N_u^4 N_{rx}^3)$, whereas the dual variables updates (step 8-10) have a complexity of $\mathcal{O}(QN_u N_s N_{RF} N_{tx} + QN_{RF}^2 N_{tx})$. Therefore, keeping only the dominant terms, the overall complexity order for the proposed algorithm is $\mathcal{O}\left(Q\left(N_{tx}^2 N_u N_s + N_{RF}^2 N_{tx}\right) + N_u^3 N_{tx} N_{rx}^2 + N_u^4 N_{rx}^3\right)$.

Table II presents the total complexity order of the proposed method and compares it against other existing low complexity alternatives namely, AM – Based [15], LASSO – Based Alt-Min (SPS and DPS) [14] and element-by-element (EBE) [20] algorithms. Taking into account that in UM-MIMO, $N_{tx}$ will tend to be very large, it means the algorithms with higher complexity will typically be EBE and the one proposed in this paper due to the terms $\mathcal{O}(QN_{tx}^2)$ and $\mathcal{O}(QN_{tx}^2 N_u N_s)$.

TABLE II
OVERALL COMPLEXITY OF DIFFERENT HYBRID PRECODING ALGORITHMS (PER SUBCARRIER).

| Operation | Complexity order |
|---|---|
| **AM – Based** | |
| Overall ([15]) | $\mathcal{O}\big(Q(N_u N_s N_{RF} N_{tx} + N_{RF}^2 N_u N_s + F^{-1} N_{RF}^3 ) + F^{-1} N_{RF}^2 N_{tx} + N_u^3 N_s^3 \big)$ |
| **LASSO – Based Alt-Min (SPS)** | |
| Overall ([14]) | $\mathcal{O}\big(Q(N_u N_s N_{RF} N_{tx} + N_{RF}^2 N_u N_s + F^{-1} N_{RF}^3 ) + N_u^2 N_s N_{RF} N_{tx} + N_u^4 N_s^3 \big)$ |
| **ADMM (from [10])** | |
| Overall ([10]) | $\mathcal{O}\big(Q(N_s N_{RF} N_{tx} + N_{RF}^2 N_{tx})\big)$ |
| **EBE** | |
| Overall ([20]) | $\mathcal{O}(QN_{tx}^2)$ |
| **Proposed** | |
| $\mathbf{F}_{RF}$ | $\mathcal{O}(QN_u N_s N_{RF} N_{tx} + F^{-1} Q N_{RF}^2 N_{tx})$ |
| $\mathbf{F}_{BB}$ | $\mathcal{O}(QN_u N_s N_{RF} N_{tx} + QN_{RF}^2 N_{tx})$ |
| $\mathbf{R}$ | $\mathcal{O}(QN_{RF} N_{tx})$ |
| $\mathbf{B}$ | $\mathcal{O}(QN_u N_s N_{RF} N_{tx})$ |
| $\mathbf{F}_{aprox}$ | $\mathcal{O}(QN_{tx}^2 N_u N_s + N_u^3 N_{tx} N_{rx}^2 + N_u^4 N_{rx}^3)$ |
| $\mathbf{U}, \mathbf{W}, \mathbf{Z}$ | $\mathcal{O}(QN_u N_s N_{RF} N_{tx} + QN_{RF}^2 N_{tx})$ |
| Overall | $\mathcal{O}\big(Q(N_{tx}^2 N_u N_s + N_{RF}^2 N_{tx}) + N_u^3 N_{tx} N_{rx}^2 + N_u^4 N_{rx}^3\big)$ |

It is important to note however, that while the computational complexity of these two design methods may be higher, both algorithms can be applied to simple AoSA/DAoSA architectures and, in particular, the proposed approach directlysupports structures with lower practical implementation complexity (and more energy efficient) such as those based on switches. Furthermore, in a single-user scenario, the interference cancellation step of the proposed algorithm is unnecessary, and the complexity reduces to $\mathcal{O}\big(Q(N_u N_s N_{RF} N_{tx} + N_{RF}^2 N_{tx})\big)$. Regarding the other algorithms, they have similar complexities. However, the AM-based algorithm is designed for single stream scenarios whereas the others consider multiuser multi-stream scenarios.

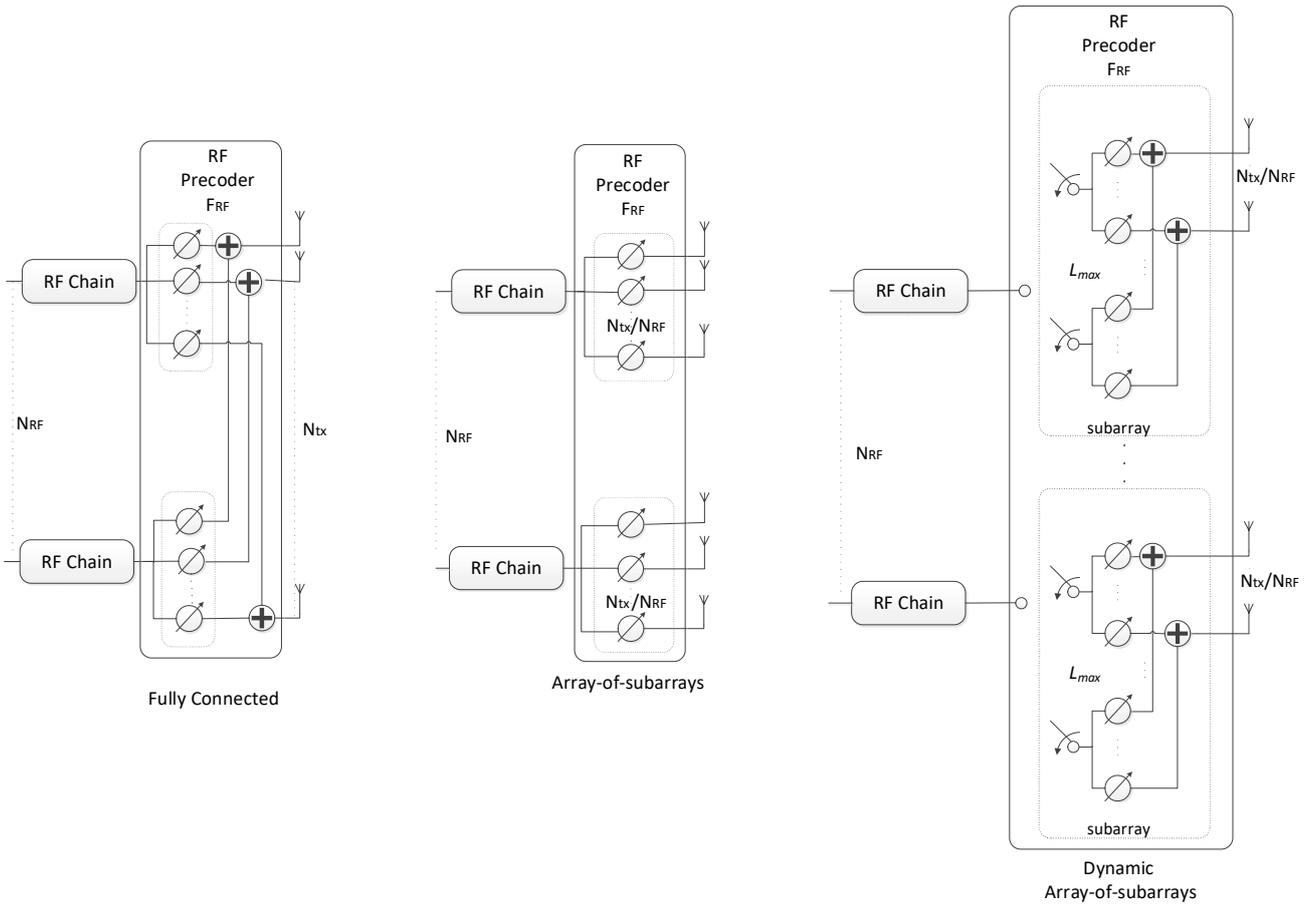

Fig. 2. Different precoder architectures for a mmWave/THz MIMO system based on phase shifters.

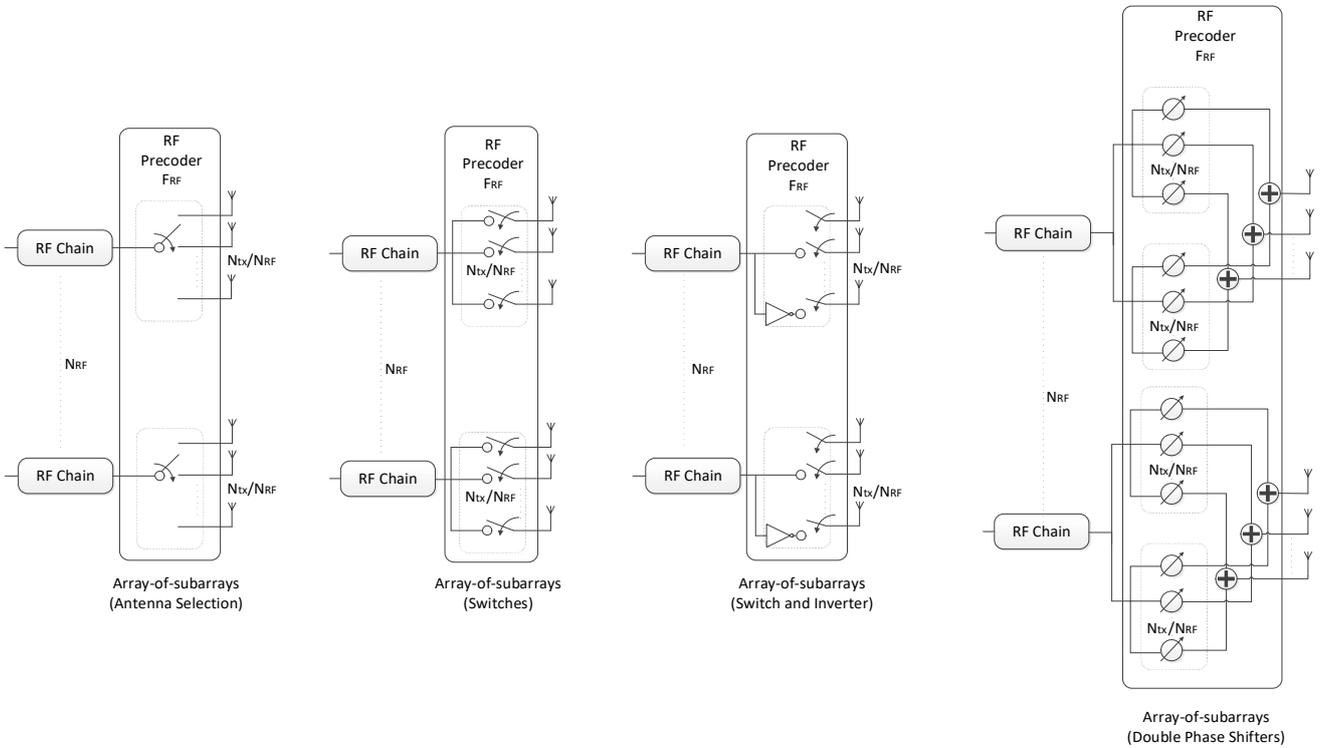

Fig. 3. Alternative implementations based on array-of-subarrays for a mmWave/THz MIMO system.

## IV. NUMERICAL RESULTS

In this section, the performance of the proposed algorithm will be evaluated and compared against other existing alternatives from the literature, considering multiuser MIMO systems. We consider that both the transmitter and receivers are equipped with uniform planar arrays (UPAs) with $\sqrt{N_{tx}} \times \sqrt{N_{tx}}$ antenna elements at the transmitter and $\sqrt{N_{rx}} \times \sqrt{N_{rx}}$ at the receiver. The respective array response vectors are given by

$$\mathbf{a}_{t/r}\left(\phi_{i,l,u}^{t/r}, \theta_{i,l,u}^{t/r}\right) = \frac{1}{\sqrt{N_{tx/rx}}} \times$$

$$\times \left[1, \ldots, e^{j\frac{2\pi}{\lambda}d\left(p\sin\phi_{i,l,u}^{t/r}\sin\theta_{i,l,u}^{t/r} + q\cos\theta_{i,l,u}^{t/r}\right)}, \right.$$

$$\left. \ldots, e^{j\frac{2\pi}{\lambda}d\left(\left(\sqrt{N_{tx/rx}}-1\right)\sin\phi_{i,l,u}^{t/r}\sin\theta_{i,l,u}^{t/r} + \left(\sqrt{N_{tx/rx}}-1\right)\cos\theta_{i,l,u}^{t/r}\right)}\right]^T, \quad (49)$$

where $p, q = 0, \ldots, \sqrt{N_{tx/rx}} - 1$ are the antenna indices, $\lambda$ is the signal wavelength and $d$ is the inter-element spacing, which we assume to be $d = \lambda/2$. We consider a sparse channel with limited scattering where $N_{\text{ray}} = 4$ and $N_{cl} = 6$. The angles of departure and arrival were selected according to a Gaussian distribution whose means are uniformly distributed in $[0, 2\pi]$ and whose angular spreads are 10 degrees. In the scenarios where we consider the existence of a LOS component, a ratio of $E\left[\left|\alpha_u^{LOS}\right|^2\right] / \sum_{i=1}^{N_{cl}}\sum_{l=1}^{N_{ray}} E\left[\left|\alpha_{i,l,u}\right|^2\right] = 10$ is assumed (in this case we are admitting very weak NLOS paths compared to LOS which is typical in the THz band [28]). A fully digital combiner was considered at each receiver and all simulation results were computed with 5000 independent Monte Carlo runs.

### A. Fully Connected Structures

First, we evaluate the performance assuming a fully connected structure. Simulation results for a scenario, where a base station with $N_{tx} = 100$ antennas transmits a single data stream ($N_s = 1$) to $N_u = 4$ users with $N_{rx} = 4$ antennas are shown in Fig. 4 for $F=1$ and Fig. 5 for $F=64$. The number of RF chains at the transmitter ($N_{RF}^{tx}$) is equal to $N_u N_s$. Besides our proposed precoder, several alternative precoding schemes are compared against the fully digital solution, namely the LASSO-Based Alt-Min, the AM-Based and ADMM-Based precoding [14], [15], [10]. It can be observed that when $F=1$, only the LASSO-Based Alt-Min with single phase shifters (SPS) and the ADMM-Based precoder from [10] (which does not remove the inter-user interference) lie far from fully digital precoder. All the others achieve near optimum results and, in fact, can even match them when adopting DPS (proposed approach and LASSO-Based Alt-Min). As explained in Section II, whereas for $F=1$ we have $\mathbf{F}_{BB}$ and $\mathbf{F}_{RF}$ designed for that specific carrier, when $F=64$, $\mathbf{F}_{RF}$ has to be common to all subcarriers. While this reduces the implementation complexity, it also results in a more demanding restriction that makes the approximation of $\mathbf{F}_{opt_k}$ (problem (5)-(7)) to become worse. Additionally, when this approximation worsens, there can also be increased interference between users. Therefore, it can be observed in the results of Fig. 5 that the gap between the fully digital precoder and all the different hybrid algorithms is substantially wider. Still, the proposed precoder manages to achieve the best results. Given the performances of the different approaches, it is important to remind that the AM-based precoding algorithm has the lowest performance in wideband but also one of the lowest computational complexity (see Table II of section III.C). In general, the proposed precoding algorithm is the one that can achieve better results at the cost of some additional computational complexity. Later on, we will address strategies based on lower complexity architectures that will allow reducing the power consumption associated to its complexity. In Fig. 6 we consider a scenario where the BS employs a larger array with $N_{tx} = 256$ antennas to transmit $N_s = 2$ simultaneous streams to each user, where $N_u = 2$. To better fit this scenario to a typical communication in the THz band we consider the existence of a LOS component, a center frequency of $f_c$=300 GHz and a bandwidth of $B$=15 GHz (it is important to note that the beam split effect is also considered in the channel model). AM precoder from [15] requires a single stream per user and thus was not included in the figure. In this scenario, the LASSO-Based Alt-Min precoding schemes present a performance substantially lower when compared to the proposed approaches. Furthermore, the best performance is achieved with the use of double phase shifters, as expected. Once again, comparing the curves of the proposed precoder against the ADMM-based precoder from [10], it is clear the advantage of adopting an interference cancellation-based design over a simple matrix approximation one.

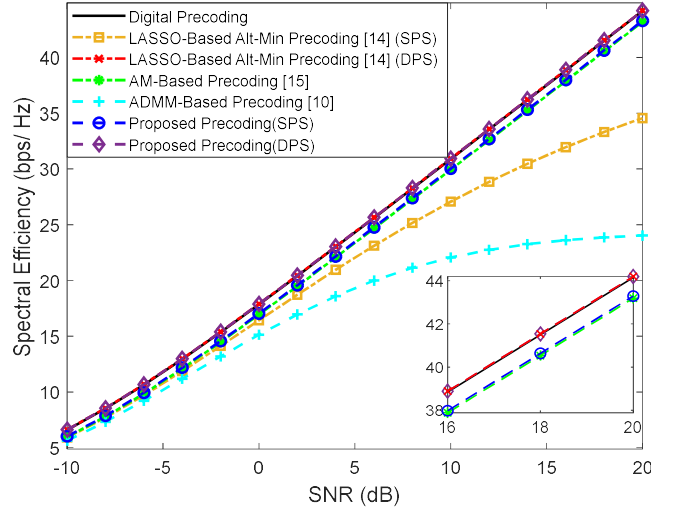

Fig. 4. Spectral efficiency versus SNR achieved by different methods with $N_u = 4$, $N_s = 1$, $N_{RF}^{tx} = 4$, $F = 1$, $N_{tx} = 100$ and $N_{rx} = 4$ (only NLOS).

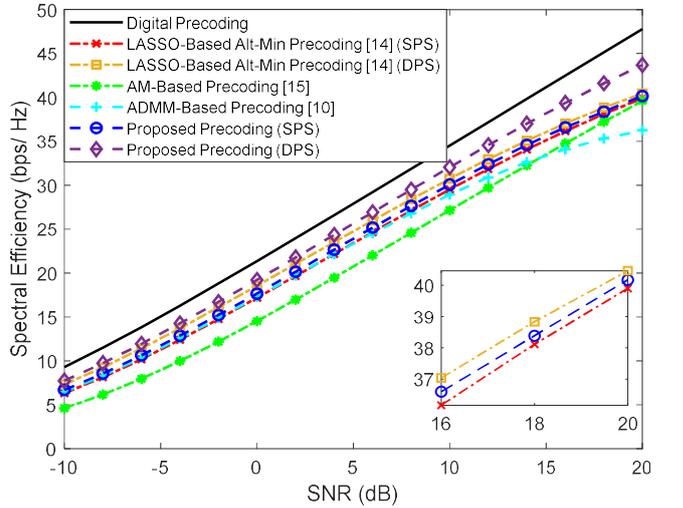

Fig. 5. Spectral efficiency versus SNR achieved by different methods with $N_u = 4$, $N_s = 1$, $N_{RF}^{tx} = 8$, $F = 64$, $N_{tx} = 100$ and $N_{rx} = 4$ (only NLOS).

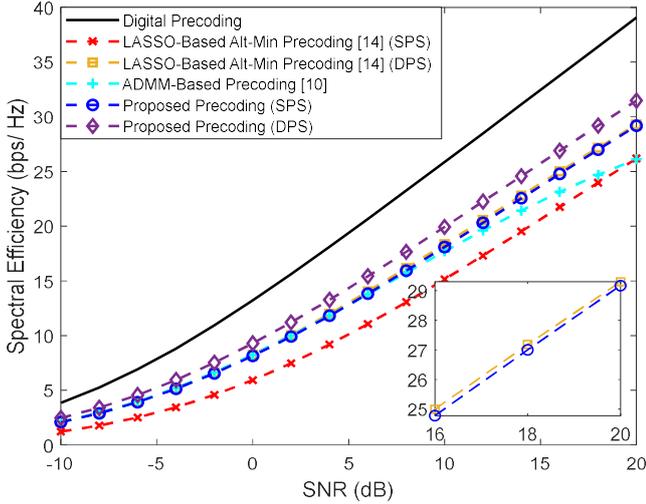
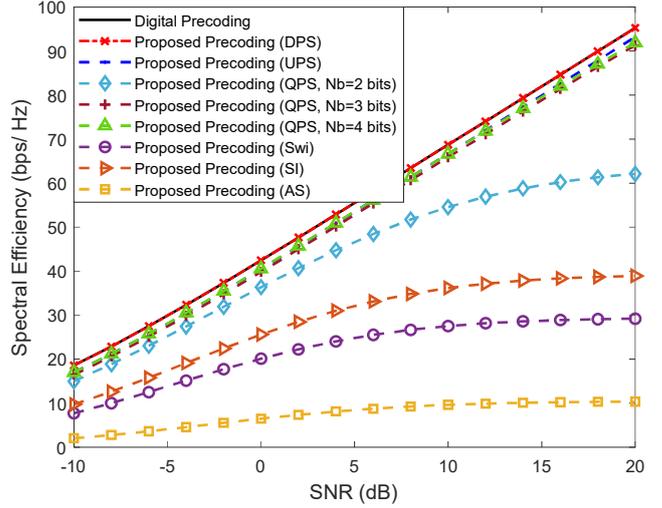

Fig. 6. Spectral efficiency versus SNR achieved by different methods with $N_u = 2$, $N_s = 2$, $N_{RF}^{tx} = 4$, $F = 64$, $N_{tx} = 256$ and $N_{rx} = 4$ (with LOS component).

Fig. 7. Spectral efficiency versus SNR achieved by the proposed precoder using different fully-connected architectures for $N_u = 4$, $N_s = 2$, $N_{RF}^{tx} = 8$, $F = 1$, $N_{tx} = 256$ and $N_{rx} = 4$ (only NLOS).

### B. Reduced Complexity Architectures

Next, we will focus on the adoption of different reduced complexity architectures according to the typologies presented in section III.B. The objective is to evaluate the performance degradation when simpler architectures are adopted.

Fig. 7 considers a scenario in which we have more than one data stream ($N_s = 2$) being sent from the BS to each user ($N_u = 4$) in a system with $N_{RF}^{tx} = N_u N_s$, $F=1$, $N_{tx} = 256$ and $N_{rx} = 4$. We considered the same penalty parameters configuration: $\rho = 0.05$, $\mu = 1$ and $\eta = \rho$. This figure is placed in a perspective of simplifying the implementation of the analog precoder but keeping a fully connected structure. We can observe that the versions based on DPS and single UPS achieve the best results, as expected. Considering the more realistic QPS versions, the results can worsen but it is visible that it is not necessary to use high resolution phase shifters since with only 3 bits resolution the results are already very close to the UPS curve. It can also be observed that the simplest of the architectures, AS, results in the worst performance but the spectral efficiency improves when the antenna selectors are replaced by a network of switches, or even better if branches with inverters are also included.

In Fig. 8, we intend to simplify the implementation even further with the adoption of AoSAs. In this case we considered that the maximum number of subarrays that can be connected to a RF chain ($L_{max}$) is only one. The scenario is the same of Fig. 7 but considers the existence of a LOS component. In fact, hereafter the existence of a LOS component is assumed for the remaining figures of the paper in order to fit the AoSA/DAoSA results to a more typical scenario in the THz band. We can observe that for AoSA structures, the degradation of the spectral efficiency is notorious, since all candidate versions present worse results when compared to the corresponding fully connected design and are all far from the fully digital solution.

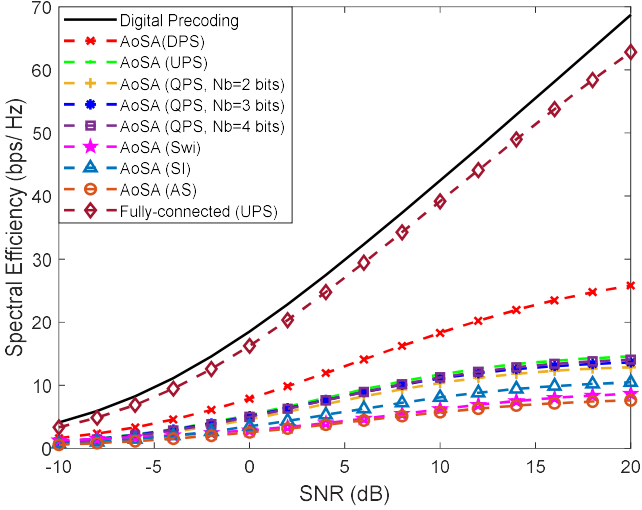

Fig. 8. Spectral efficiency versus SNR achieved by the proposed precoder using different AoSA architectures with $L_{\max}=1$, $N_u=4$, $N_s=2$, $N_{RF}^{tx}=8$, $F=1$, $N_{tx}=256$ and $N_{rx}=4$ (with LOS component).

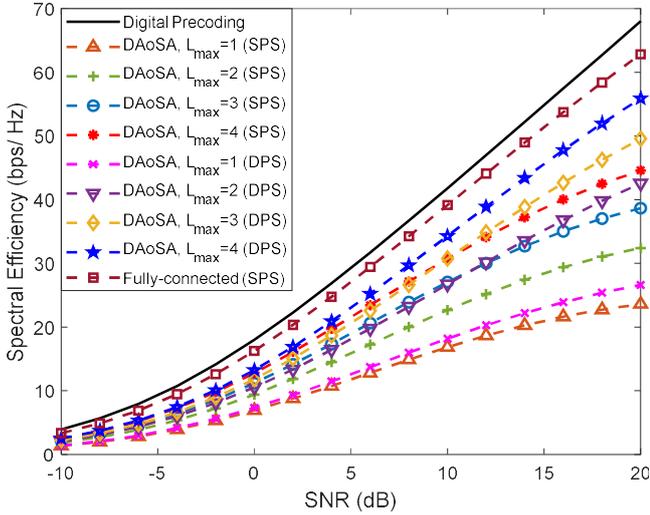

Fig. 9. Spectral efficiency versus SNR achieved by the proposed precoder considering an architecture based on DAoSAs and the variation of the maximum number of subarrays that can be connected to a RF chain ($L_{\max}$) for $N_u=4$, $N_s=2$, $N_{RF}^{tx}=8$, $F=1$, $N_{tx}=256$ and $N_{rx}=4$ (with LOS component).

To reduce the large performance loss due to the adoption of a simple AoSA architecture, we can allow the dynamic connection of more subarrays to each RF chain by adopting a DAoSA structure, as introduced in section III.B. In Fig. 9 we study the effect increasing the maximum number of subarrays that can be connected to an RF chain ($L_{\max}$) in the performance of these schemes. Each subarray has a size of 32 antennas ($n_t$). Curves assuming the use of SPS as well as of DPS are included. It can be observed that the increase in the number of connections to subarrays, $L_{\max}$, has a dramatic effect on the performance, resulting in a huge improvement by simply going from $L_{max}=1$ to $L_{max}=2$. Increasing further to $L_{max}=4$, the results become close to the fully connected case showing that the DAoSA can be a very appealing approach for balancing the spectral efficiency with hardware complexity and power consumption. Combining the increase of $L_{max}$ with the adoption of DPS can also improve the results but the gains become less pronounced for $L_{max}>1$. It is important to note that the penalty parameters can be fine-tuned for different system configurations. One of the objectives of adopting these low complexity solutions is to reduce the overall power consumption. Based on [20], we can calculate the total power consumption of each precoding scheme using

$$P_C = P_{BB}N_{BB} + (P_{DAC}+P_{OS}+P_M)N_{RF}^{tx} + P_{PA}N_{tx} + \\ +P_{PC}N_{tx}+P_{PS}N_{PS}+P_{SWI}N_{SWI}+P_T \quad (50)$$

where $P_{BB}$ is the power of the baseband block (with $N_{BB}=1$), $P_{DAC}$ is the power of a DAC, $P_{OS}$ is the power of an oscillator, $P_M$ is the power of a mixer, $P_{PA}$ is the power of a power amplifier, $P_{PC}$ is the power of a power combiner, $P_{PS}$ is the power of a phase shifter, $P_{SWI}$ is the power of a switch and $P_T$ denotes the transmit power. The $N_x$ variable represents the number of elements of each device used in the precoder configuration.

Based on the values provided in [20] and [39] for the power consumption of individual devices in the 300 GHz band we adopt the following values: $P_{BB}$=200 mW, $P_{DAC}$=110 mW, $P_{OS}$=4 mW, $P_M$=22 mW, $P_{PA}$=60 mW, $P_{PC}$=6.6 mW, $P_{SWI}$=24 mW and $P_T$=100 mW. Regarding the phase shifters, we assume values of $P_{PS}$=10, 20, 40, 100 mW for 1, 2, 3 and 4 quantization bits. Considering the same configuration scenario as Figures 7-9 with $N_u=4$, $N_s=2$, $N_{RF}^{tx}=N_uN_s$, $F=1$, $N_{tx}=256$ and $N_{rx}=4$, we provide the values of power consumption for different precoder configurations in Table III.

For the fully-connected structure with UPS, we assumed that $P_{PS}$=100 mW which corresponds to quantized phase shifter with $N_b$=4 bits [39]. For the remaining phase-shifter based precoder structures we assumed that $P_{PS}$=40 mW which corresponds to quantized phase shifters with $N_b$=3 bits, since with only 3 bits resolution the results are already very close to the UPS curve (see Fig. 7). As can be seen from this table, the use of architectures based on DAoSAs allows us to reduce considerably the amount of power that is consumed at the precoder. In fact, we can reduce up to 55% the amount of consumed power if we consider a precoder scheme based on DAoSA with DPS and $L_{max}$=4 versus a FC structure precoder based on UPS, with only a small performance penalty (Fig. 9). This saving increases to 73% if the DPS structure is replaced by an SPS one. In the particular case of architectures based on quantized phase shifters, we observed that by decreasing the number of quantization bits, it is possible to substantially reduce the power consumption without excessively compromising the complexity (as seen in Fig. 7). The conclusion is corroborated by [20] and [39], since the architectures based on low resolution QPS, AoSAs and DAoSAs present a superior energy efficiency when compared to the fully-connected structure with UPS.

TABLE III
POWER CONSUMPTION FOR DIFFERENT IMPLEMENTATIONS OF THE
PROPOSED PRECODER FOR $N_u = 4$, $N_s = 2$, $N_{RF}^{tx} = 8$, $F = 1$ AND $N_{tx} = 256$.

| | Precoder | Estimated power consumption [W] |
|---|---|---|
| Fully-Connected | DPS | 428.04 |
| | UPS | 223.24 |
| | QPS ($N_b$=2) | 59.4 |
| | QPS ($N_b$=3) | 100.36 |
| | SWI | 67.59 |
| | SI | 38.92 |
| DAoSA SPS | $L_{max}$=1 | 28.87 |
| | $L_{max}$=2 | 39.30 |
| | $L_{max}$=3 | 49.73 |
| | $L_{max}$=4 | 60.17 |
| DAoSA DPS | $L_{max}$=1 | 39.11 |
| | $L_{max}$=2 | 59.78 |
| | $L_{max}$=3 | 80.45 |
| | $L_{max}$=4 | 101.13 |

In Fig. 10 and Fig. 11, we provide a comparison between our proposed precoder and the EBE precoder from [20], considering an architecture based on DAoSAs (with SPS) and a scenario configuration similar to Fig. 9, i.e., with $N_s = 2$, $N_{RF}^{tx} = 8$, $F = 1$, $N_{tx} = 256$ and $N_{rx} = 4$. These figures present various curves where the maximum number of subarrays that can be connected to a RF chain, $L_{max}$, is changed. Fig. 10 refers to a SU scenario ($N_u$=1) whereas Fig. 11 corresponds to a MU scenario with $N_u$=4. In the SU case, the proposed precoder achieves results very close to the fully digital precoder, even with only $L_{max}$=2. Compared to the proposed algorithm, EBE shows a wider gap, even though it has smaller complexity (as presented in Table II of section III.C). When we increase the number of users from $N_u = 1$ to $N_u = 4$, we can clearly observe that the EBE algorithm suffers a substantial degradation compared to the proposed solution which can be explained due to the lack of inter-user interference cancellation (it was not specifically designed for MU scenarios).

Even though a sub-6 GHz system often adopts fully digital processing [40], where each antenna element has a dedicated RF chain, it is possible to apply the proposed hybrid design algorithm to a sub-6GHz channel since it is independent of a specific MIMO channel (as are the other alternative algorithms that we used as benchmarks and which are targeted at solving the matrix approximation problem).

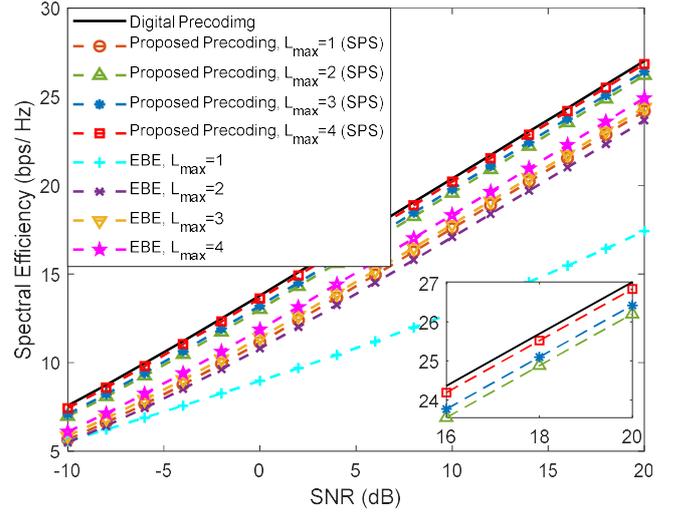

Fig. 10. Spectral efficiency versus SNR achieved by the proposed precoder and by the EBE algorithm considering an architecture based on DAoSAs and the variation of the maximum number of subarrays that can be connected to a RF chain ($L_{max}$) for $N_u = 1$, $N_s = 2$, $N_{RF}^{tx} = 8$, $F = 1$, $N_{tx} = 256$ and $N_{rx} = 4$ (with LOS component).

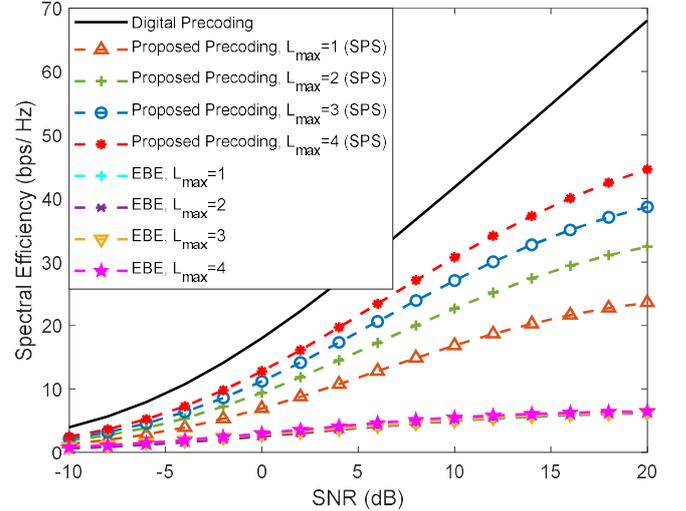

Fig. 11. Spectral efficiency versus SNR achieved by the proposed precoder and by the EBE algorithm considering an architecture based on DAoSAs and the variation of the maximum number of subarrays that can be connected to a RF chain ($L_{max}$) for a mmWave/THz system with $N_u = 4$, $N_s = 2$, $N_{RF}^{tx} = 8$, $F = 1$, $N_{tx} = 256$ and $N_{rx} = 4$ (with LOS component).

To exemplify, Fig. 12 presents the simulated results obtained for the same scenario of Fig. 4 but considering an ideal uncorrelated channel which approximates a rich scattering environment that is typical in sub-6 GHz bands. It can be observed that the proposed approach displays similar behavior to the ones in the upper-bands channel, showing that it can also be used for this particular type of channels (even though it may require a higher number of RF chains to achieve a good approximation to the fully digital solution in some scenarios, due to the channel not being sparse, as noted in [41]). While we have shown how the proposed approach can deal with several relevant types of analog precoders/combiners, it is important to note that are other alternative structures that have been recently proposed in the literature. For example, some authors have considered precoding paradigms based on time-delayers structures for THz systems [28], [42]. One of the most notorious is the Delay Phase Precoding (DPD), which consists in the use of a Time Delay (TD) network between the RF chains and the traditional phase shifters network in order to convert phase-controlled analog precoding into delay-phase controlled analog precoding. The main advantage related with this type of precoding is that the time delays in the TD network are carefully designed to generate frequency-dependent beams which are aligned with the spatial directions over the whole bandwidth [42]. While we do not address the adoption of time-delay structures in this paper, it should be possible do derive a projection algorithm that simultaneously takes into account the constraints imposed in both analog-precoding steps: time-delay network and frequency-independent phase-shifters.

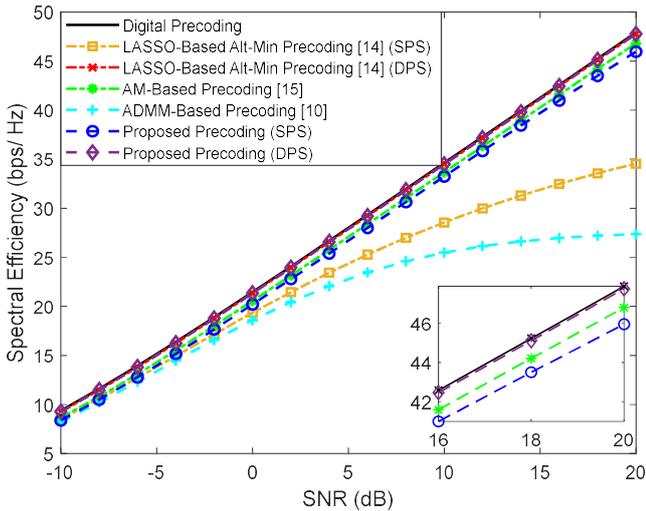

Fig. 12. Spectral efficiency versus SNR achieved by different methods for a mmWave/THz MIMO-OFDM system with $N_u = 4$, $N_s = 1$, $N_{RF}^{tx} = 4$, $F = 1$, $N_{tx} = 100$ and $N_{rx} = 4$ considering an uncorrelated channel.

## V. CONCLUSION

In this paper, we proposed an iterative algorithm for hybrid precoding design which is suitable for multiuser MIMO systems operating in mmWave and THz bands. The adopted approach splits the formulated design into a sequence of smaller subproblems with closed-form solutions and can work with a broad range of configuration of antennas, RF chains and data streams. The separability of the design process allows the adaptability of the algorithm to different architectures, making it suitable to be implemented with low-complexity AoSA and DAoSA structures which are particularly relevant for the deployment of ultra-massive MIMO in hardware constrained THz systems. It was shown that good trade-offs between spectral efficiency and hardware implementation complexity can in fact be achieved by the proposed algorithm for several different architectures.

ACKNOWLEDGMENT

This work was supported by the FCT - Fundação para a Ciência e Tecnologia under the grant 2020.05621.BD. The authors also acknowledge the funding provided by FCT/MCTES through national funds and when applicable co-funded EU funds under the project UIDB/50008/2020.

## REFERENCES

[1] T.S. Rappaport, et al., "Millimeter wave mobile communications for 5G cellular: It will work!", *IEEE Access*, vol. 1, pp. 335–349, 2013.

[2] A. Uwaechia and N. Mahyuddin, "A Comprehensive Survey on Millimeter Wave Communications for Fifth-Generation Wireless Networks: Feasibility and Challenges", *IEEE Access*, vol. 8, pp. 62367-62414, 2020.

[3] Akyildiz, A. Kak and S. Nie, "6G and Beyond: The Future of Wireless Communications Systems", *IEEE Access*, vol. 8, pp. 133995-134030, 2020.

[4] J. Tan and T. Dai, "THz Precoding for 6G: Applications, Challenges, Solutions, and Opportunities", arXiv preprint, arXiv:2005.10752, May 2020.

[5] H. Sarieddeen, M. Alouini, T. Al-Naffouri, "An Overview of Signal Processing Techniques for Terahertz Communications", arXiv preprint, arXiv:2005.13176, May 2020.

[6] C. Lin and G. Li, "Terahertz Communications: An Array-of-Subarrays Solution", *IEEE Communications Magazine*, vol. 54, no. 12, pp. 124-131, 2016.

[7] I. Ahmed, H. Khammari, A. Shahid, A. Musa, K.S. Kim, E. De Poorter, I. Moerman, "A survey on hybrid beamforming techniques in 5G: Architecture and system model perspectives", *IEEE Commun. Surv. Tutor.*, Vol. 20, no. 4, pp. 3060–3097, 2018.

[8] R. Chataut and R. Akl, "Massive MIMO Systems for 5G and beyond Networks—Overview, Recent Trends, Challenges, and Future Research Direction", *Sensors*, vol. 20, no. 10, pp. 1-35, 2020.

[9] F. Sohrabi, W. Yu, "Hybrid digital and analog beamforming design for large-scale antenna arrays", *IEEE J. Sel. Top. Signal Process.*, vol. 10, no. 3, pp. 501–513, 2016.

[10] N. Souto, J. Silva, J. Pavia and M. Ribeiro, "An alternating direction algorithm for hybrid precoding and combining in millimeter wave MIMO systems", Physical Communication, vol. 34, pp. 165-173, 2019.

[11] J. P. Pavia, N. Souto, M. Ribeiro, J. Silva and R. Dinis, "Hybrid Precoding and Combining Algorithm for Reduced Complexity and Power Consumption Architectures in mmWave Communications", The 2020 IEEE 91st Vehicular Technology Conference: VTC2020-Spring, 2020.

[12] K. Guan, G. Li, T. Kurner, A.F. Molisch, B. Peng, R. He, B. Hui, J. Kim, Zhangdui Zhong, "On millimeter wave and THz mobile radio channel for smart rail mobility", *IEEE Trans. Veh. Techn.*, vol. 66, no. 7, pp. 5658–5674, 2017.

[13] A. Alkhateeb, J. Mo, N. Gonzáles-Prelcic, R.W. Heath, Jr, "MIMO Precoding and combining solutions for millimeter-wave systems", *IEEE Commun. Mag.*, vol.52, no.12, pp.122–131, 2014.

[14] X. Yu, J. Zhang and K. Letaief, "Alternating minimization for hybrid precoding in multiuser OFDM mmWave systems", 2016 50th Asilomar Conference on Signals, Systems and Computers, 2016.

[15] H. Yuan, J. An, N. Yang, K. Yang and T. Duong, "Low Complexity Hybrid Precoding for Multiuser Millimeter Wave Systems Over Frequency Selective Channels", *IEEE Transactions on Vehicular Technology*, vol. 68, no. 1, pp. 983-987, 2019.

[16] F. Liu, X. Kan, X. Bai, R. Du, H. Liu and Y. Zhang, "Hybrid precoding based on adaptive RF-chain-to-antenna connection for millimeter

wave MIMO systems", *Physical Communication*, vol. 39, pp. 1-9, 2020.

[17] Xu, Y. Cai, M. Zhao, Y. Niu and L. Hanzo, "MIMO-Aided Nonlinear Hybrid Transceiver Design for Multiuser mmWave Systems Relying on Tomlinson-Harashima Precoding", arXiv preprint, arXiv:2008.05860, August 2020.

[18] A. Vizziello, P. Savazzi and K. Chowdhury, "A Kalman Based Hybrid Precoding for Multi-User Millimeter Wave MIMO Systems", IEEE Access, vol. 6, pp. 55712-55722, 2018.

[19] H. M. Elmagzoub, "On the MMSE-based multiuser millimeter wave MIMO hybrid precoding design", *International Journal of Communication Systems*, vol. 33, no. 11, pp. 1-17, 2020.

[20] L. Yan, C. Han and J. Yuan, "A Dynamic Array-of-Subarrays Architecture and Hybrid Precoding Algorithms for Terahertz Wireless Communications", *IEEE Journal on Selected Areas in Communications*, pp. 1-16, 2020.

[21] S. Busari, K. Huq, S. Mumtaz and J. Rodriguez, "Terahertz Massive MIMO for Beyond-5G Wireless Communication", ICC 2019 - 2019 IEEE International Conference on Communications (ICC), 2019.

[22] S. Boyd, N. Parikh, E. Chu, B. Peleato, J. Eckstein, Distributed optimization and statistical learning via the alternating direction method of multipliers, Found. Trends Mach. Learn. 3 (1) (2011) 1–122.

[23] Y. Wang, W. Yin, J. Zeng, "Global convergence of ADMM in nonconvex nonsmooth optimization", Journal of Scientific Computing, vol. 78, pp. 29-63, 2019.

[24] X. Yu, J. Shen, J. Zhang and K. B. Letaief, "Alternating Minimization Algorithms for Hybrid Precoding in Millimeter Wave MIMO Systems," in IEEE Journal of Selected Topics in Signal Processing, vol. 10, no. 3, pp. 485-500, April 2016, doi: 10.1109/JSTSP.2016.2523903.

[25] A. Alkhateeb and R. W. Heath, "Frequency Selective Hybrid Precoding for Limited Feedback Millimeter Wave Systems," in IEEE Transactions on Communications, vol. 64, no. 5, pp. 1801-1818, May 2016, doi: 10.1109/TCOMM.2016.2549517.

[26] H. Yuan, N. Yang, K. Yang, C. Han and J. An, "Hybrid Beamforming for Terahertz Multi-Carrier Systems Over Frequency Selective Fading," in IEEE Transactions on Communications, vol. 68, no. 10, pp. 6186-6199, Oct. 2020, doi: 10.1109/TCOMM.2020.3008699.

[27] S. Tarboush, H. Sarieddeen, H. Chen, M. H. Loukil, H. Jemaa, M. S. Alouini, T. Y. Al-Naffouri, "TeraMIMO: A Channel Simulator for Wideband Ultra-Massive MIMO Terahertz Communications," arXiv preprint, arXiv:2104.11054, Apr. 2021.

[28] C. Lin, G. Y. Li and L. Wang, "Subarray-Based Coordinated Beamforming Training for mmWave and Sub-THz Communications," *IEEE Journal on Selected Areas in Communications*, vol. 35, no. 9, pp. 2115-2126, 2017.

[29] D. Nguyen, L. Le, T. Le-Ngoc and R. Heath, "Hybrid MMSE Precoding and Combining Designs for mmWave Multiuser Systems", IEEE Access, vol. 5, pp. 19167-19181, 2017.

[30] D. Bertsekas, Nonlinear programming. Belmont, Massachusetts: Athena Scientific, 2016.

[31] O. El Ayach, S. Rajagopal, S. Abu-Surra, Z. Pi, and R. W. Heath, "Spatially sparse precoding in millimeter wave MIMO systems", IEEE Transactions on Wireless Communications, vol. 13, no. 3, pp. 1499–1513, 2014.

[32] X. Yu, J. Shen, J. Zhang and K. Letaief, "Alternating Minimization Algorithms for Hybrid Precoding in Millimeter Wave MIMO Systems", IEEE Journal of Selected Topics in Signal Processing, vol. 10, no. 3, pp. 485-500, 2016.

[33] Q. H. Spencer, A.L. Swindlehurst, M. Haardt, "Zero-forcing methods for downlink spatial multiplexing in multiuser MIMO channels", IEEE Transactions on Signal Processing, vol.52, no.2, pp.461 - 471, 2004.

[34] R. Mendez-Rial, C. Rusu, N. Gonzalez-Prelcic, A. Alkhateeb and R. Heath, "Hybrid MIMO Architectures for Millimeter WaveCommunications: Phase Shifters or Switches?", IEEE Access, vol. 4, pp. 247-267, 2016

[35] J. Lee and Y. H. Lee, "AF relaying for millimeter wave communication systems with hybrid RF/baseband MIMO processing," 2014 IEEE International Conference on Communications (ICC), Sydney, NSW, pp.5838-5842, 2014.

[36] S. Payami, M. Ghoraishi, M. Dianati and M. Sellathurai, "Hybrid Beamforming with a Reduced Number of Phase Shifters for Massive MIMO Systems", IEEE Transactions on Vehicular Technology, vol. 67, no. 6, pp. 4843-4851, 2018.

[37] M. Tian, J. Zhang, Y. Zhao, L. Yuan, J. Yang and G. Gui, "Switch and Inverter Based Hybrid Precoding Algorithm for mmWave Massive MIMO System: Analysis on Sum-Rate and Energy-Efficiency", IEEE Access, vol. 7, pp. 49448-49455, 2019.

[38] X. Yu, J. Zhang and K.B. Letaief, "Doubling Phase Shifters for Efficient Hybrid Precoder Design in Millimeter-Wave Communication Systems", Journal of Communications and Information Networks, vol. 4, pp. 51-67, 2019.

[39] L. Yan, C.Han, N. Yang and J. Yuan, "Dynamic-subarray with Quantized- and Fixed-phase Shifters for Terahertz Hybrid Beamforming", GLOBECOM 2020 - 2020 IEEE Global Communications Conference, 2020.

[40] Z. Li, C. Zhang, I. Lu and X. Jia, "Hybrid Precoding Using Out-of-Band Spatial Information for Multi-User Multi-RF-Chain Millimeter Wave Systems", IEEE Access, vol. 8, pp. 50872-50883, 2020. Available: 10.1109/access.2020.2979712.

[41] S. Park, A. Alkhateeb and R. W. Heath, "Dynamic Subarrays for Hybrid Precoding in Wideband mmWave MIMO Systems," in IEEE Transactions on Wireless Communications, vol. 16, no. 5, pp. 2907-2920, May 2017, doi: 10.1109/TWC.2017.2671869.

[42] J. Tan and L. Dai, "Delay-phase precoding for THz massive MIMO with beam split," in Proc. IEEE Global Commun. Conf., Hawaii, USA, 2019.